\documentclass[1p]{elsarticle}
\usepackage[T1]{fontenc}
\usepackage[utf8]{inputenc}
\usepackage{geometry}
\usepackage{graphicx}
\usepackage{esint}


\usepackage{hyperref}
\usepackage{babel}
\begin{document}
\begin{frontmatter}
\title{The Dynamics of Ions on Phased Radio-frequency Carpets in High Pressure
Gases and Application for Barium Tagging in Xenon Gas Time Projection
Chambers}
\author[3]{B.J.P.~Jones\corref{cor}}
\author[3]{A.~Raymond}
\author[3]{K.~Woodruff\corref{cor}}
\author[3]{N.~Byrnes}
\author[4]{A.A.~Denisenko}
\author[4]{F.W.~Foss}
\author[3]{K.~Navarro}
\fntext[c]{NEXT Co-spokesperson.}
\author[3]{D.R.~Nygren\fnref{c}}
\author[4]{T.T.~Vuong}
\author[2]{C.~Adams}
\author[11]{H.~Almaz\'an}
\author[26]{V.~\'Alvarez}
\author[24]{B.~Aparicio}
\author[22]{A.I.~Aranburu}
\author[7]{L.~Arazi}
\author[20]{I.J.~Arnquist}
\author[16]{S.~Ayet}
\author[5]{C.D.R.~Azevedo}
\author[2]{K.~Bailey}
\author[26]{F.~Ballester}
\author[21]{J.M.~Benlloch-Rodr\'{i}guez}
\author[14]{F.I.G.M.~Borges}
\author[11]{S.~Bounasser}
\author[19]{S.~C\'arcel}
\author[19]{J.V.~Carri\'on}
\author[27]{S.~Cebri\'an}
\author[20]{E.~Church}
\author[14]{C.A.N.~Conde}
\author[11]{T.~Contreras}
\author[21,9]{F.P.~Coss\'io}
\author[25]{G.~D\'iaz}
\author[19]{J.~D\'iaz}
\author[16]{T.~Dickel}
\author[14]{J.~Escada}
\author[26]{R.~Esteve}
\author[11]{A.~Fahs}
\author[7]{R.~Felkai}
\author[13]{L.M.P.~Fernandes}
\author[21,9]{P.~Ferrario}
\author[5]{A.L.~Ferreira}
\author[13]{E.D.C.~Freitas}
\author[22,9]{Z~Freixa}
\author[21]{J.~Generowicz}
\author[8]{A.~Goldschmidt}
\author[21,9]{J.J.~G\'omez-Cadenas\fnref{c}}
\author[21]{R.~Gonz\'alez}
\author[25]{D.~Gonz\'alez-D\'iaz}
\author[11]{R.~Guenette}
\author[10]{R.M.~Guti\'errez}
\author[11]{J.~Haefner}
\author[2]{K.~Hafidi}
\author[1]{J.~Hauptman}
\author[13]{C.A.O.~Henriques}
\author[25]{J.A.~Hernando~Morata}
\author[21,23]{P.~Herrero-G\'omez}
\author[26]{V.~Herrero}
\author[11]{J.~Ho}
\author[7]{Y.~Ifergan}
\author[25]{M.~Kekic}
\author[18]{L.~Labarga}
\author[3]{A.~Laing}
\author[6]{P.~Lebrun}
\author[11]{D.~Lopez Gutierrez}
\author[26]{N.~L\'opez-March}
\author[10]{M.~Losada}
\author[13]{R.D.P.~Mano}
\author[19]{J.~Mart\'in-Albo}
\author[19]{A.~Mart\'inez}
\author[7]{G.~Mart\'inez-Lema}
\author[21,19]{M.~Mart\'inez-Vara}
\author[3]{A.D.~McDonald}
\author[2]{Z.E.~Meziani}
\author[3]{K.~Mistry}
\author[21,9]{F.~Monrabal}
\author[13]{C.M.B.~Monteiro}
\author[26]{F.J.~Mora}
\author[19]{J.~Mu\~noz Vidal}
\author[19]{P.~Novella}
\author[21]{E.~Oblak}
\author[21]{M.~Odriozola-Gimeno}
\author[25,19]{B.~Palmeiro}
\author[6]{A.~Para}
\author[12]{J.~P\'erez}
\author[19]{M.~Querol}
\author[7]{A.B.~Redwine}
\author[25]{J.~Renner}
\author[17]{L.~Ripoll}
\author[21,9]{I.~Rivilla}
\author[10]{Y.~Rodr\'iguez Garc\'ia}
\author[26]{J.~Rodr\'iguez}
\author[23]{C.~Rogero}
\author[3]{L.~Rogers}
\author[21,12]{B.~Romeo}
\author[19]{C.~Romo-Luque}
\author[14]{F.P.~Santos}
\author[13]{J.M.F. dos~Santos}
\author[7]{A.~Sim\'on}
\author[19]{M.~Sorel}
\author[11]{C.~Stanford}
\author[13]{J.M.R.~Teixeira}
\author[4]{P.~Thapa}
\author[26]{J.F.~Toledo}
\author[21]{J.~Torrent}
\author[19]{A.~Us\'on}
\author[5]{J.F.C.A.~Veloso}
\author[15]{R.~Webb}
\fntext[e]{On leave from Soreq Nuclear Research Center, Yavneh, Israel.}
\author[7]{R.~Weiss-Babai\fnref{e}}
\fntext[f]{Deceased.}
\author[15]{J.T.~White\fnref{f}}
\author[19]{N.~Yahlali}
\address[1]{
Department of Physics and Astronomy, Iowa State University, Ames, IA 50011-3160, USA}
\address[2]{
Argonne National Laboratory, Argonne, IL 60439, USA}
\address[3]{
Department of Physics, University of Texas at Arlington, Arlington, TX 76019, USA}
\address[4]{
Department of Chemistry and Biochemistry, University of Texas at Arlington, Arlington, TX 76019, USA}
\address[5]{
Institute of Nanostructures, Nanomodelling and Nanofabrication (i3N), Universidade de Aveiro, Campus de Santiago, Aveiro, 3810-193, Portugal}
\address[6]{
Fermi National Accelerator Laboratory, Batavia, IL 60510, USA}
\address[7]{
Nuclear Engineering Unit, Faculty of Engineering Sciences, Ben-Gurion University of the Negev, P.O.B. 653, Beer-Sheva, 8410501, Israel}
\address[8]{
Lawrence Berkeley National Laboratory (LBNL), 1 Cyclotron Road, Berkeley, CA 94720, USA}
\address[9]{
Ikerbasque (Basque Foundation for Science), Bilbao, E-48009, Spain}
\address[10]{
Centro de Investigaci\'on en Ciencias B\'asicas y Aplicadas, Universidad Antonio Nari\~no, Sede Circunvalar, Carretera 3 Este No.\ 47 A-15, Bogot\'a, Colombia}
\address[11]{
Department of Physics, Harvard University, Cambridge, MA 02138, USA}
\address[12]{
Laboratorio Subterr\'aneo de Canfranc, Paseo de los Ayerbe s/n, Canfranc Estaci\'on, E-22880, Spain}
\address[13]{
LIBPhys, Physics Department, University of Coimbra, Rua Larga, Coimbra, 3004-516, Portugal}
\address[14]{
LIP, Department of Physics, University of Coimbra, Coimbra, 3004-516, Portugal}
\address[15]{
Department of Physics and Astronomy, Texas A\&M University, College Station, TX 77843-4242, USA}
\address[16]{
II. Physikalisches Institut, Justus-Liebig-Universitat Giessen, Giessen, Germany}
\address[17]{
Escola Polit\`ecnica Superior, Universitat de Girona, Av.~Montilivi, s/n, Girona, E-17071, Spain}
\address[18]{
Departamento de F\'isica Te\'orica, Universidad Aut\'onoma de Madrid, Campus de Cantoblanco, Madrid, E-28049, Spain}
\address[19]{
Instituto de F\'isica Corpuscular (IFIC), CSIC \& Universitat de Val\`encia, Calle Catedr\'atico Jos\'e Beltr\'an, 2, Paterna, E-46980, Spain}
\address[20]{
Pacific Northwest National Laboratory (PNNL), Richland, WA 99352, USA}
\address[21]{
Donostia International Physics Center, BERC Basque Excellence Research Centre, Manuel de Lardizabal 4, San Sebasti\'an / Donostia, E-20018, Spain}
\address[22]{
Department of Applied Chemistry, Universidad del Pais Vasco (UPV/EHU), Manuel de Lardizabal 3, San Sebasti\'an / Donostia, E-20018, Spain}
\address[23]{
Centro de F\'isica de Materiales (CFM), CSIC \& Universidad del Pais Vasco (UPV/EHU), Manuel de Lardizabal 5, San Sebasti\'an / Donostia, E-20018, Spain}
\address[24]{
Department of Organic Chemistry I, University of the Basque Country (UPV/EHU), Centro de Innovaci\'on en Qu\'imica Avanzada (ORFEO-CINQA), San Sebasti\'an / Donostia, E-20018, Spain}
\address[25]{
Instituto Gallego de F\'isica de Altas Energ\'ias, Univ.\ de Santiago de Compostela, Campus sur, R\'ua Xos\'e Mar\'ia Su\'arez N\'u\~nez, s/n, Santiago de Compostela, E-15782, Spain}
\address[26]{
Instituto de Instrumentaci\'on para Imagen Molecular (I3M), Centro Mixto CSIC - Universitat Polit\`ecnica de Val\`encia, Camino de Vera s/n, Valencia, E-46022, Spain}
\address[27]{
Centro de Astropart\'iculas y F\'isica de Altas Energ\'ias (CAPA), Universidad de Zaragoza, Calle Pedro Cerbuna, 12, Zaragoza, E-50009, Spain}
\cortext[cor]{Corresponding Authors}

\begin{abstract}
Radio-frequency (RF) carpets with ultra-fine pitches are examined for ion transport in gases at atmospheric pressures and above. We develop new analytic and computational methods for modeling RF ion transport at densities where dynamics are strongly influenced by buffer gas collisions. An analytic description of levitating and sweeping forces from phased arrays is obtained, then thermodynamic and kinetic principles are used to calculate ion loss rates in the presence of collisions. This methodology is validated against detailed microscopic SIMION simulations. We then explore a parameter space of special interest for neutrinoless double beta decay experiments: transport of barium ions in xenon at pressures from 1 to 10 bar.   Our computations account for molecular ion formation and pressure dependent mobility as well as finite temperature effects.   We discuss the challenges associated with achieving suitable operating conditions, which lie beyond the capabilities of existing devices, using presently available or near-future manufacturing techniques.
\end{abstract}
\end{frontmatter}

\section{Introduction}

\begin{figure}[t]
\begin{centering}
\includegraphics[width=0.99\columnwidth]{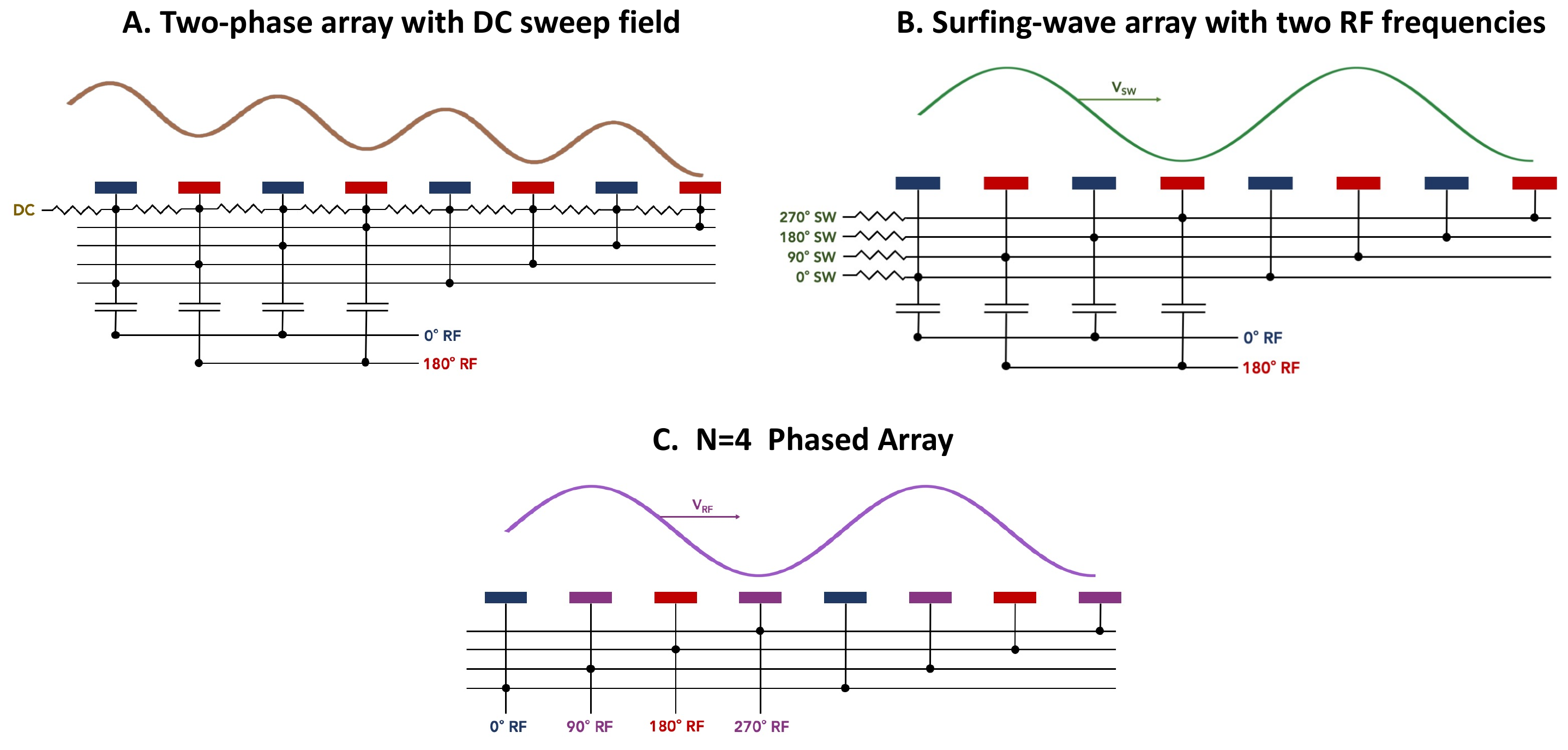}
\par\end{centering}
\caption{Diagrams showing the drive circuits of three RF carpet configurations:
A) two-phase array with DC sweep field, as considered in Ref.~\cite{schwarz2011rf};
B) superposed two-phase levitation with $N$-phase surfing wave, as discussed
in Ref.~\cite{bollen2011ion}; C) $N$-phased RF array, the subject
of the present work.\label{fig:Diagrams-showing-the}}
\end{figure}

Radio-frequency (RF) carpets and funnels are structures used in mass
spectrometry~\cite{kim2000design,shaffer1997novel} and gas-phase
stopper cells for nuclear physics~\cite{schury2016status,savard2008radioactive} that
apply a rapidly switching voltage to generate ion levitation and transport
in gaseous media. The dynamics of RF levitation
are similar to those of a Paul ion trap~\cite{paul1990electromagnetic}.
In such devices, the applied voltage generates a micro-motion of
the trapped ion at the RF frequency. Since the field is non-uniform,
at different points in the micro-motion the ion experiences a different
strength of electric field. At sufficiently low buffer gas densities
the correlation between the electric field strength and position
 in the micro-cycle generates a time-integrated effective
force, even though the time-integrated electric field is
zero at every point in space. This force can be associated with a pseudo-potential, as first
described by Dehmelt (and is reviewed briefly in the Appendix).    A DC ``push field'' superposed onto
the repulsive RF pseudo-potential can then be used to generate a narrow ion trapping region above the carpet surface,
along which ions can be transported laterally.
The Dehmelt treatment can be applied directly to predict the pseudo-potential
for a two-phase RF carpet (shown in Fig.~\ref{fig:Diagrams-showing-the},
A) since the electric field is sinusoidal in time at every position
in space. This calculation has been presented by Schwarz~\cite{schwarz2011rf},
alongside predictions of the corresponding regions of stability for
two-phase RF carpets.   

Building upon this underlying principle, RF carpets with more complex
dynamics have been proposed. Particularly notable is the introduction
by Bollen of the ion-surfing scheme~\cite{bollen2011ion}.
Here, two superposed AC signals are applied across a phased electrode
structure. First, a fast RF voltage is applied between adjacent electrode
pairs to generate a levitating force via Dehmelt potential; second,
a sweeping phased wave at a different frequency is applied over the two-phase RF voltage. This
system has ion dynamics that are intractable analytically, and can
involve both phase-locked and tumbling ion motions, depending on the
surfing frequency. An example of a circuit to realize this system
is shown in Fig.~\ref{fig:Diagrams-showing-the}, B. Simulations of
this geometry were performed to explore the regions of stability and
dynamics in Ref.~\cite{bollen2011ion}, which have a nontrivial structure.
Demonstrations~\cite{arai201456,brodeur2013experimental,gehring2016221,querci2018rf,ranjan2011new}
of ion dynamics on various RF carpet devices based on these modalities
have been presented for pressures of up 300 millibar, usually in helium
buffer gas. 

In this paper we present computational studies of an RF carpet operating
mode based on a purely $N$-phased RF voltage. This device has superficial similarity to the surfing
wave system of Bollen, but with a single high frequency phased RF
wave responsible for both the levitation and translational dynamics.
This operating mode is shown in Fig. \ref{fig:Diagrams-showing-the},
C. In this mode the ion trajectories are analytically calculable,
and we will show that analytic formulae match well the results detailed
microscopic simulations in the widely used SIMION simulation package~\cite{appelhans2005simion}. Once augmented with stochastic effects,
these calculations provide an excellent description of the microscopically simulated
RF carpet performance. In this operating mode the same RF wave is responsible for both levitation
and transport, simplifying the theoretical treatment and allowing for an analytical solution to the ion motion.

The application that motivates this work is the search for neutrinoless double beta decay in gas-phase $^{136}$Xe detectors.
In such experiments, efficient capture
and identification of single barium ions or ``barium tagging''~\cite{moe:1991ik}, is widely understood
to be a technology that could offer major advances in experimental
sensitivity though dramatic reduction of backgrounds from radiogenic
and cosmogenic activity. Barium tagging is an area of intense R\&D
within the neutrinoless double beta decay community~\cite{rivilla2020fluorescent,bainglass2018mobility,Cesar2014,jones2016single,mcdonald2018demonstration,mong2015spectroscopy,nEXOSingle,nygren2015detecting,sinclair2011prospects,thapa2019barium,thapa2021demonstration,twelker2015barium,brunner2015rf}, and several techniques have now been demonstrated
for single barium ion imaging in xenon~\cite{flatt:2007aa,mcdonald2018demonstration,nEXOSingle}. Ion collection and extraction from a ton to multi-ton scale volume
of liquid or high pressure gas remains major unsolved problem, however
~\cite{brunner2015rf,twelker2014apparatus}.
RF-based methods for this purpose have been explored in other contexts. Notably, Ref.~\cite{brunner2015rf} 
proposes to employ gas
flow to drive ions through an RF-only funnel for extraction from 10
bar pressure to a vacuum ion trap for atomic fluorescence imaging.
Here we explore an alternate approach, the application of ultra-fine pitch RF carpets
to the problem of ion transport in dense xenon gas for
barium tagging.  The required carpets are of similar form to those proposed for ion catching in atmospheric pressure air in Ref~\cite{poteshin2020investigation}.
In this scenario, RF transport is implemented
without gas flow to sweep barium ions arriving at a detector cathode
to a few small single molecule fluorescence sensors~\cite{thapa2021demonstration}.  A key goal of this work is to evaluate what operating parameters would need to 
be achieved to allow for RF transport in high pressure xenon gas experiments,
and assess the viability of achieving these beyond-state-of-the-art parameters
 with available or near-term manufacturing techniques.

The first part of this paper is general in scope, and serves to develop our formalism. 
 Sec.~\ref{sec:Dynamics-of-ion}
presents analytic calculations of the micro- and macro-trajectories
of ion motion on $N$-phased carpets. We demonstrate that the macro-motion
can be considered as driven by a Dehmelt-esque pseudo-potential in the perpendicular
direction with a steady sweep force in the transverse. We provide
analytic expressions for the levitating force, equilibrium ion height,
transverse velocity and micro-motion radius. Sec.~\ref{sec:Finite-electrode-width}
compares this calculation, which has idealized point-like electrodes,
to a finite element simulation of the electric field and microscopic
simulation of smooth ion motion in that geometry. This serves both as a validation 
of the analytic treatment in the far-field region and a calculation
 of the distance scales where near-field finite electrode width effects
become relevant to ion transport.  We find that the analytic
calculation is accurate for our purposes at distances above around
0.25 times the RF carpet pitch $p$, whereas near-field
distortions to the electric field due to surface geometry will cause
ion losses below 0.25$\times p$. Sec.~\ref{sec:Stochastic-effects-from}
introduces stochastic effects from collisions. We develop a new thermodynamic
approach to predicting ion distributions above RF carpets, and
show that the random motion of ions due to collisions can be understood
as a thermalization to the buffer gas temperature within the pseudo-potential well.
We then provide a kinetic theory derivation of the expected ion loss
rates from the tails of this potential into the RF carpet surface.
Both predictions are validated against computationally intensive collision-by-collision
simulations in SIMION~\cite{appelhans2005simion}.  Sec.~\ref{sec:Comparative-analysis-of}
 explores the transport efficiency of different
RF carpet phasings constrained by either fixed peak-to-peak voltage
or fixed maximum electrode-to-electrode voltage. In the latter practically
important scenario, carpets with 4-6 phases are found to be the most
effective.

Having developed the appropriate suite of analytic techniques we proceed to delineate the required operating
parameters of RF carpets for barium collection in high pressure xenon gas.  Section~\ref{sec:GeneralConsideration} 
discusses the problem in general terms, and briefly outlines application specifications
 in terms of geometry, coverage, power and pressure.  In Sec.~\ref{sec:Operating-Parameters-for}
we establish the key RF carpet operating parameters for high pressure
xenon gas environments, paying careful attention to the details of ion transport microphysics 
in this environment. Sec.~\ref{sec:Operating-conditions-for} presents
our main result, an assessment of the required carpet parameters for
transport barium ions in high pressure xenon gas neutrinoless double
beta decay experiments. Sec.~\ref{sec:Conclusions} presents our
conclusions, which are that existing carpet structures based on precision
printed circuit boards (PCBs)~\cite{arai201456} appear viable at pressures
up to one bar, whereas 10~bar operation for the neutrinoless double beta decay application
 exceeds the capabilities of such structures.  The required operating parameters for use at 10 bar may potentially
  be accessible with devices based on micro-fabrication techniques
used for micro-electromechanical systems and trapped ion quantum information science applications~\cite{sterling2014fabrication,brown2021materials,mehta2014ion} and large wafer CMOS processes.
We briefly discuss the near-term R\&D needed to realize such devices.

\section{Dynamics of ion levitation and transport on phased RF carpets: analytic
treatment \label{sec:Dynamics-of-ion}}

\begin{figure}[t]
\begin{centering}
\includegraphics[width=0.49\columnwidth]{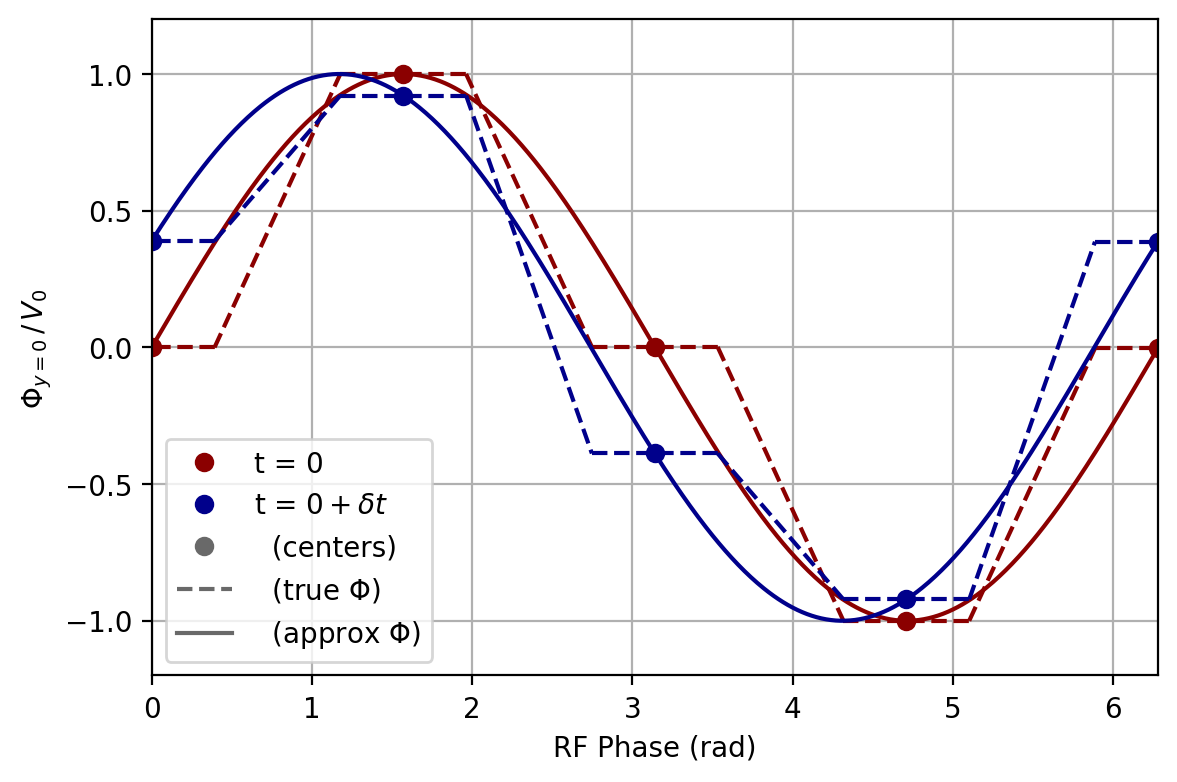}\includegraphics[width=0.49\columnwidth]{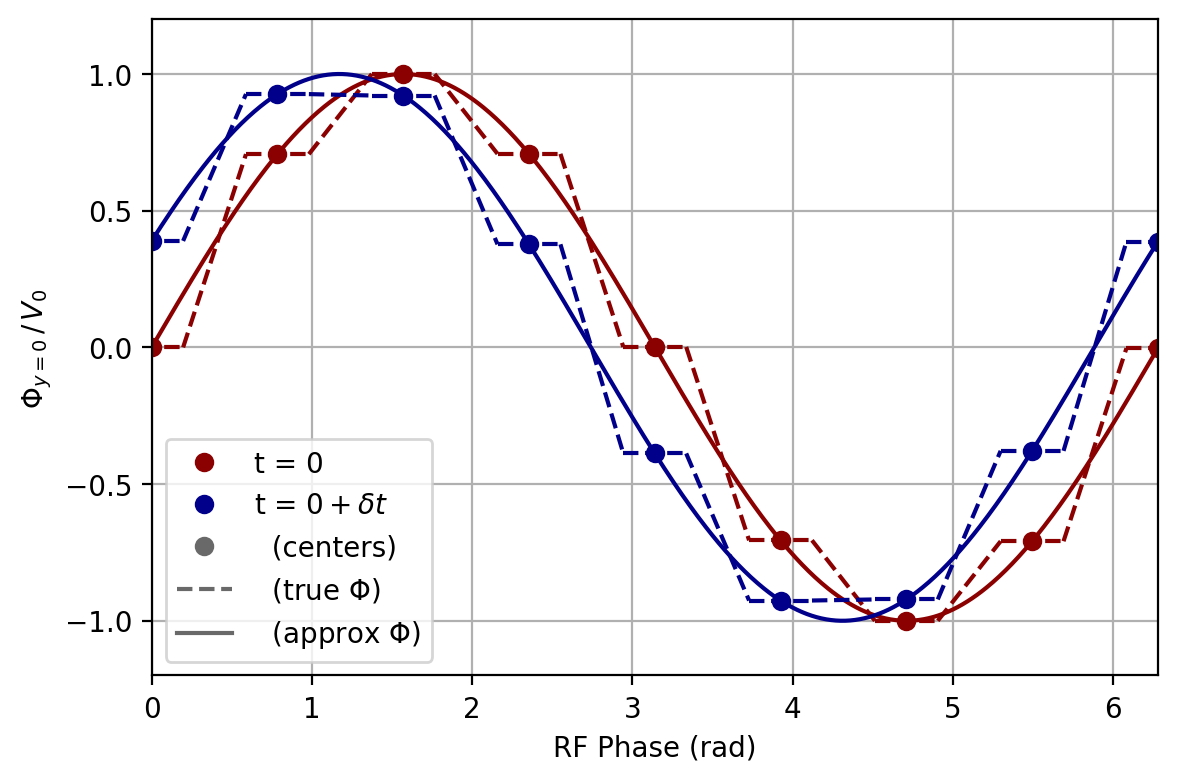}
\par\end{centering}
\caption{Smooth potential approximation along carpet surface for a four-phase
array (left) and eight-phase array (right) \label{fig:Smooth-potential-approximation}}
\end{figure}

\begin{figure}[t]
\begin{centering}
\includegraphics[width=0.5\columnwidth]{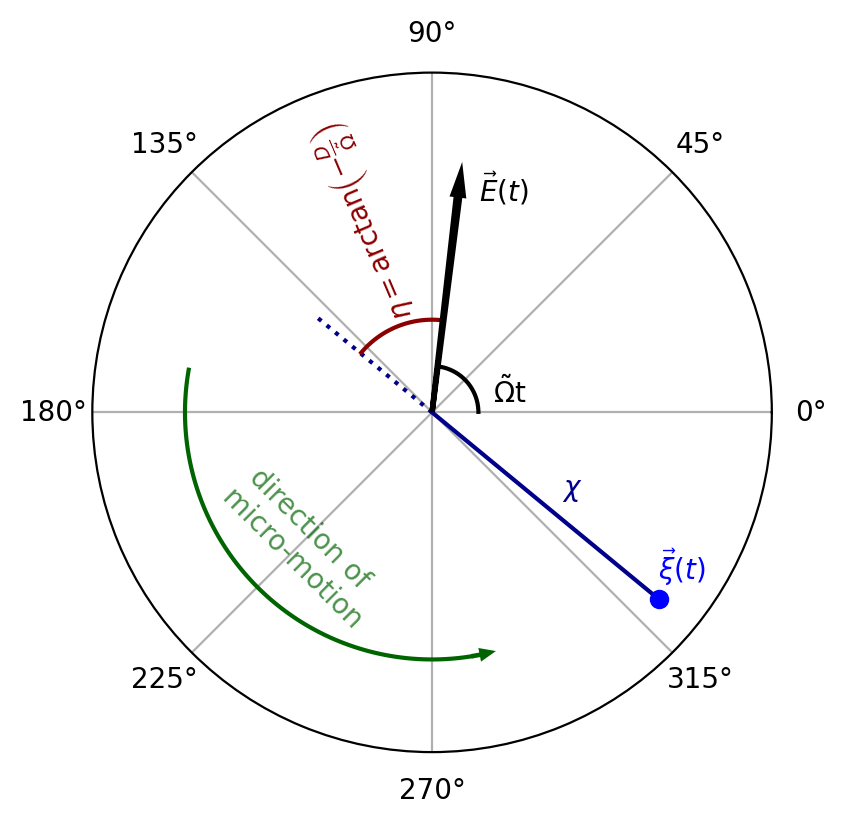}\includegraphics[width=0.5\columnwidth]{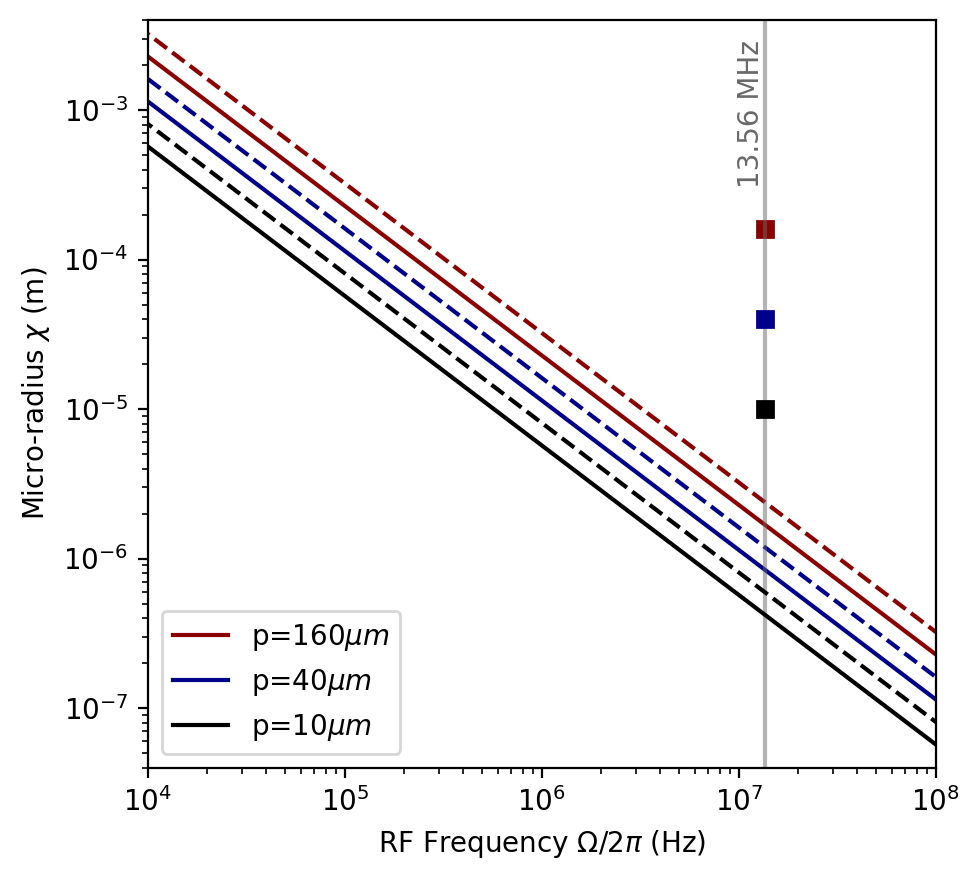}
\par\end{centering}
\caption{Left: Polar plot showing relationship between $E$-field direction and micro-motion vector in terms of damping constant $D$ and effective local frequency $\tilde{\Omega}$. Right: Comparison of predicted micro-radius for N=2 (solid) and N=4 (dashed) lines for $E_{push}=20$~V cm$^{-1}$ and m=952 amu (a representative cluster mass of a barium dication in xenon gas at 10 bar) vs. frequency.  For the operating parameter point considered later in this work ($\Omega$=13.56 MHz) the micro-radius is always much smaller than the carpet pitch (shown as square markers).\label{fig:RFIllustration}}

\end{figure}

A purely $N$-phased RF carpet powered by a peak-to-peak RF voltage of
$V_{pp}$ is one with voltage $V_{j}$ on the $j$th electrode at
time $t$ of:
\begin{equation}
V_{j}(t)=\frac{V_{pp}}{2}\sin\left(\Omega t+\frac{2\pi j}{N}\right).
\end{equation}
For the purposes of our calculations we make the approximation that
this generates a smooth potential along the carpet surface of the
form:
\begin{equation}
\Phi(x,y,t)|_{y=0}=\frac{V_{pp}}{2}\sin\left(\Omega t+\frac{2\pi}{Np}x\right),\label{PotSurface}
\end{equation}
where we are using a coordinate system such that $x$ runs radially along the carpet
surface and $y$ runs perpendicularly to it, with $y=0$ defining the carpet surface.
This approximation neglects the finite width of the electrodes and
the flat potential across them, as shown in Fig.~\ref{fig:Smooth-potential-approximation}.
The approximation becomes increasingly precise for larger the values
of $N$, and we will explore its limitations quantitatively in Sec.
\ref{sec:Finite-electrode-width}.
Using the potential Eq.~\ref{PotSurface} as a boundary condition
for Laplace's equation the electric potential over all space can be
obtained:
\begin{equation}
\Phi_{0}(x,y,t)=\frac{V_{pp}}{2}\sin\left(\Omega t+\frac{2\pi}{Np}x\right)\exp\left(-\frac{2\pi}{Np}y\right),
\end{equation}
from which the driving electric field can be found. It is of a fixed magnitude at every position, rotating at constant
angular velocity:
\begin{equation}
\vec{E}(\vec{r},t)=\frac{V_{pp}}{2}\frac{2\pi}{Np}\exp\left(-\frac{2\pi}{Np}y\right)\vec{u}\left[\Omega t+\frac{2\pi}{Np}x\right],\quad\quad\quad\vec{u}\left(\theta\right)=\left(\begin{array}{c}
\cos\left(\theta\right)\\
-\sin\left(\theta\right)\\
0
\end{array}\right).
\end{equation}
Notably this is not of Dehmelt form typical in a Paul trap, which
would require an electric field like $E(x,t)=E(x)e^{i\Omega t}$;
nevertheless, we will see similar trapping dynamics in the vertical
direction emerge. We seek solutions to the following driven equation
of motion, where the ion moves in response to both the applied RF
voltages and a superimposed DC push field $E_{push}$ toward the carpet:
\begin{equation}
m\ddot{\vec{r}}+\frac{q}{\mu}\dot{\vec{r}}=q\left[\vec{E}(\vec{r},t)-E_{push}\hat{y}\right].\label{eq:EOM}
\end{equation}
As is conventional with treatment of radio-frequency trap systems
we make an assumption that the trajectory $\vec{r}(t)$ can be factorized
into a micro-motion $\vec{\xi}(t)$ and macro-motion $\vec{z}(t)$:
\begin{equation}
\vec{r}(t)=\vec{z}(t)+\vec{\xi}(t),\label{eq:Separation}
\end{equation}
where the dynamics of $\vec{\xi}$ are determined by the RF frequency
and the dynamics of $\vec{z}$ are slower and emergent. Unlike the
simple Dehmelt-like trap, in this system the relevant frequency of
the micro-motion is not exactly $\Omega$, because the ion is moving
laterally while passing through the RF field. It is convenient to
introduce ion co-moving coordinates as $\vec{x}=\vec{r}-\langle \vec{v}(y)\rangle t$
where $\langle \vec{v}(y)\rangle$ is the mean ion velocity at fixed height $y$;
in these coordinates we find that
\begin{equation}
\vec{E}(x,t)=\frac{V_{pp}}{2}\frac{2\pi}{Np}\exp\left(-\frac{2\pi}{Np}y\right)\vec{u}\left[\tilde{\Omega}(y)t\right],\quad\quad\quad\tilde{\Omega}(y)=\left(\Omega+\frac{2\pi}{Np}\langle \vec{v}(y)\rangle\right).\label{eq:EField}
\end{equation}
For transverse trajectories at fixed $\langle \vec{v}\rangle$, the ion will experience
an electric field which is rotating spatially at frequency $\tilde{\Omega}$,
and this is thus the frequency of the micro-motion. We will need to
solve for $\langle \vec{v}\rangle$ generated by this micro-motion, which is also
required to predict $\tilde{\Omega}$ and thus the trajectory mean
height, in an iterative manner. Substitution of the electric field
into the equation of motion and neglecting the slow-varying derivatives
of macro-coordinates yields an equation of motion for the micro-motion:
\begin{equation}
m\ddot{\vec{\xi}}+\frac{q}{\mu}\dot{\vec{\xi}}=q\frac{V_{pp}}{2}\frac{2\pi}{Np}\exp\left(-\frac{2\pi}{Np}y\right)\left(\begin{array}{c}
\cos\tilde{\Omega}t\\
-\sin\tilde{\Omega}t\\
0
\end{array}\right),\label{eq:EOfM-1-4-1}
\end{equation}
with y being the y-component of the macro-coordinate $\vec{z}$. Eq.~\ref{eq:EOfM-1-4-1} can be solved for $\vec{\xi}(t):$
\begin{equation}
\vec{\xi}(t)=-\chi\,\vec{u}[\tilde{\Omega}t-\eta],\quad\quad\chi=\frac{q}{m\tilde{\Omega}\sqrt{D^{2}+\tilde{\Omega}^{2}}}\frac{V_{pp}}{2}\frac{2\pi}{Np}\exp\left(-\frac{2\pi}{Np}y\right).
\end{equation}
The micro-motion executes circles of radius $\chi$ around the smoothly
moving $\vec{z}(t)$ macro-trajectory. We have introduced the
constant $D$ associated with damping of the ballistic ion motion
by gas collisions and the phase factor $\eta$ which dictates by how
much the micro-motion phase trails the E-field direction:
\begin{equation}
D=\frac{q}{\mu m},\quad\mathrm{and}\quad\eta=\arctan\left[-D/\tilde{\Omega}\right].
\end{equation}
The relationship between the directions of $\vec{\xi}(t)$ and $\vec{E}(t)$ are illustrated in Fig~\ref{fig:RFIllustration}, left. To solve for the macro-motion, we insert Eq.~\ref{eq:Separation}
into Eq.~\ref{eq:EOM}, the equation of motion, and average over one
full micro-cycle. We find the governing equation for the macro-motion
$\vec{z}(t)$:
\begin{equation}
m{\langle\vec{\ddot{z}}}\rangle_{\tilde{\Omega}}+\frac{q}{\mu}\langle\dot{\vec{z}}\rangle_{\tilde{\Omega}}=q\langle\left(\vec{\xi}.\vec{\nabla}\right)\vec{E}[\vec{z}]\rangle_{\tilde{\Omega}}-qE_{push}\hat{y},\label{eq:EOfM-1-5-1-1}
\end{equation}
into which we can substitute $\vec{\xi}$ from Eq.~\ref{eq:EOfM-1-4-1} and extract the effective force driving $\vec{z}(t)$
emerging from the micro-motion:
\begin{equation}
F_{i}=-\frac{q^{2}}{m\tilde{\Omega}\sqrt{D^{2}+\tilde{\Omega}^{2}}}\langle\nabla_{j}E_{i}[\vec{z}]R_{jk}[-\eta]E_{k}\rangle_{\tilde{\Omega}},
\end{equation}
following methods analogous to those outlined in the Appendix. Here $R_{jk}[\theta]$ is a rotation matrix in the xy plane by angle
$\theta$. Evaluating this effective force explicitly:
\begin{equation}
\vec{F}=\frac{q^{2}}{m\tilde{\Omega}\sqrt{\left(D^{2}+\tilde{\Omega}^{2}\right)}}\left(\frac{2\pi}{Np}\right)^{3}\left(\frac{V_{pp}}{2}\right)^{2}\exp\left(-\frac{4\pi}{Np}y\right)\left(\begin{array}{c}
\sin\left(\eta\right)\\
\cos\left(\eta\right)
\end{array}\right).
\end{equation}
Note that $\eta=0$ is the fully ballistic case; in this scenario
we find a force which acts only in $y$ to repel the ion from the
carpet. On the other hand as $\eta$ becomes larger a force pushing
the ion along the carpet also appears. Once the motion becomes completely
viscous, the force levitating the ion above the carpet disappears
entirely. Dependence of the phase $\eta$ upon the relative contributions
of ballistic vs.\ damped dynamics thus dictates the extent to which
the micro-motion conspires with the electric field to provide a levitating
force. 

The ion transverse velocity $\langle v_x\rangle$ at height $y$ can be found by considering
the transverse component of the 2D equation of motion, averaged over
a micro-cycle. In terms of the $x$-component of the effective force $F_x$:
\begin{equation}
\langle v_x\rangle=\frac{\mu}{q}F_{x}=\frac{\mu q}{m\tilde{\Omega}\sqrt{\left(D^{2}+\tilde{\Omega}^{2}\right)}}\left(\frac{2\pi}{Np}\right)^{3}\left(\frac{V_{pp}}{2}\right)^{2}\exp\left(-\frac{4\pi}{Np}y\right)\sin\left(\eta\right)
\end{equation}
\begin{equation}
=\frac{Dq\mu}{m\tilde{\Omega}\left(D^{2}+\tilde{\Omega}^{2}\right)}\left(\frac{2\pi}{Np}\right)^{3}\left(\frac{V_{pp}}{2}\right)^{2}\exp\left(-\frac{4\pi}{Np}y\right).\label{eq:TransVel}
\end{equation}
The final equality requires application of trigonometric identity
to obtain $\sin\left[\eta\right]=-\frac{D}{\sqrt{D^{2}+\Omega^{2}}}$.
If we assume that $\delta=\frac{2\pi\langle \vec{v}\rangle}{Np\Omega}$ is a small
parameter so that it is appropriate to use $\Omega$ rather than $\tilde{\Omega}$
in the expression for $v$, we find the leading order approximation:
\begin{equation}
v\sim v_{0}=\frac{Dq\mu}{m\Omega\left(D^{2}+\Omega^{2}\right)}\left(\frac{2\pi}{Np}\right)^{3}\left(\frac{V_{pp}}{2}\right)^{2}\exp\left(-\frac{4\pi}{Np}y\right).\label{eq:TransVel-1}
\end{equation}
In any situation where this approximation is insufficient, it is possible
to proceed iteratively to calculate higher order corrections to $\vec{v}$
if necessary. For example, the first order correction to $\langle \vec{v} \rangle$
can be found by substituting $\langle \vec{v}\rangle_{0}$ into $\tilde{\Omega}$
in Eq.~\ref{eq:TransVel} and expanding to first order in $\delta$:
\begin{equation}
v=v_{0}\left[1-\frac{2\pi}{Np\Omega}\frac{D^{2}+3\Omega^{2}}{D^{2}+\Omega^{2}}+...\right].
\end{equation}
We will not generally need these higher order terms in this work, the leading order
expression being accurate with only percent-level corrections.
This completes the treatment of the macro-dynamics in the horizontal
direction. In the vertical direction, the ion will be acted on by
two competing forces, the repulsive force generated by the micro-motion
and the attractive force generated by the push field. These two forces
balance when
\begin{equation}
qE_{push}-\frac{q^{2}}{m\tilde{\Omega}\sqrt{\left(D^{2}+\tilde{\Omega}^{2}\right)}}\left(\frac{2\pi}{Np}\right)^{3}\left(\frac{V_{pp}}{2}\right)^{2}\exp\left(-\frac{4\pi}{Np}y\right)\cos\left(\eta\right)=0.
\end{equation}
We can thus consider the system as being trapped within a Dehmelt-like
pseudo-potential of the form
\begin{equation}
V=\frac{q^{2}}{m\left(D^{2}+\tilde{\Omega}^{2}\right)}\frac{1}{2}\left(\frac{2\pi}{Np}\right)^{2}\left(\frac{V_{pp}}{2}\right)^{2}\exp\left(-\frac{4\pi}{Np}y\right)+qE_{push}y.\label{eq:EffPot}
\end{equation}
Where once again we have employed a useful trigonometric identity
yielding $\cos\left[\eta\right]=\frac{\Omega}{\sqrt{D^{2}+\Omega^{2}}}$.
The pseudo-potential has a minimum at height $\langle y\rangle$ given by
\begin{equation}
\langle y\rangle=\frac{Np}{4\pi}\ln\left[\frac{2qV_{pp}^{2}}{m\left(D^{2}+\tilde{\Omega}^{2}\right)E_{push}}\left(\frac{\pi}{Np}\right)^{3}\right],
\end{equation}
which is the height at which the ion levitates above the carpet for
stable transverse motion.  The radius of the micro-motion
at this height has an especially simple form, which is independent
of both ion mobility and RF driving voltage,
\begin{equation}
\chi(\langle y\rangle)=\sqrt{\frac{Np}{2\pi}\frac{qE_{push}}{m\tilde{\Omega^{2}}}},
\end{equation}
For a stable trajectory to exist, we must enforce $\chi\ll\langle y\rangle$, otherwise the micro-trajectory will intercept the carpet surface during its motion. Fig.~\ref{fig:RFIllustration}, right shows the calculated micro-radii for the parameter points considered later in this work, and shows that the micro-radii at the driving frequency of interest will always be small relative to the carpet pitch, and much smaller than the hence stable trajectory height.  The micro-radius need not be considered further in what follows.

The transverse velocity of the ion at this height also has an especially
simple form, and is to leading order:
\begin{equation}
\langle v_x\rangle=\mu E_{push}\frac{D}{\Omega}.\label{eq:TransVel-1-1}
\end{equation}
The effective vertical trap depth in electronvolts can be found by
simply evaluating \textbf{$V(\langle y\rangle)-V(0)$}. The trap depth has
dependencies on all of $N,p,\mu,m,q,V_{pp}$ and $E_{push}$ via Eq.
\ref{eq:EffPot}, and further exploration follows in later sections.

\section{Finite electrode width effects beyond the leading order geometrical
approximation \label{sec:Finite-electrode-width}}

In the calculations of the two-phase RF carpet performed by Schwarz~\cite{schwarz2011rf}, the Dehmelt potential is calculated accounting
for the effects of finite electrode width and assuming linear voltage
drop across the intermediate insulator. The calculations for the phased
arrays presented above neglect these effects and assume a sinusoidal
profile over the carpet surface. This approximation is expected to
become increasingly good as $N$ becomes large. We can gain an estimate
of the magnitude of these effects for the worst case $N=2$ by comparing
the effective Dehmelt potential from Schwarz with that for an equivalent $N=2$
phased array treated under our approximation scheme.  In both cases, the expression for
the pseudo-potential in the $y$ direction reduces to the form:
\begin{equation}
V_{N=2}(y)=E_{push}y+X\frac{1}{\tilde{\Omega}^{2}+D^{2}}\frac{q}{m}\left(\frac{V_{pp}}{p}\right)^{2}\exp(-2\pi y/p).
\end{equation}
The pre-factor $X$ depends on which calculation is used. For
the Schwartz-like calculation with gap-to-pitch ratio of 0.5, we find $X=\frac{8}{\pi^{2}}$,
and for gap-to-pitch ratio of 0 it is $X=1$. For our $N$-phase calculation
with $N=2$ which assumes a sinusoidal surface potential (realized approximately but not exactly
for all gap-to-pitch ratios), we obtain $X=\frac{\pi^{2}}{8}$. 
Since the effective
levitating force is proportional to $V_{pp}^{2}$, taking these three
results as an ensemble we can conclude that the detailed structure
of the electrodes for a given pitch maps to an uncertainty on the required RF voltage of around
$\pm8.5\%$, loosely estimated as the standard deviation of the three approximation schemes.
\begin{figure}[t]
\begin{centering}
\includegraphics[width=0.99\columnwidth]{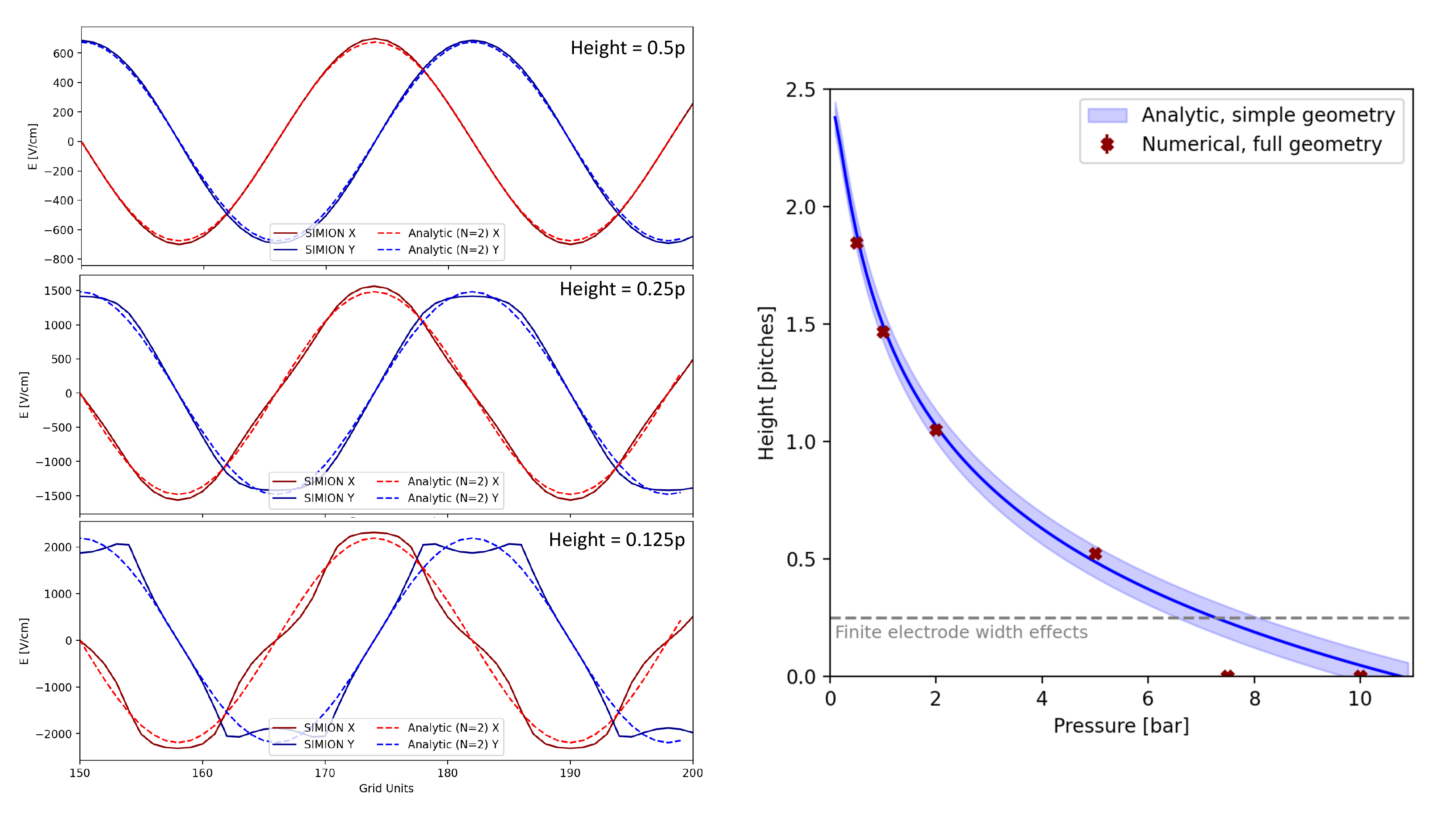}
\par\end{centering}
\caption{Finite electrode width effects on RF carpet transport. Left: comparison
of idealized calculation of electric fields to full numerical simulation
of RF carpet geometry, evaluating both X and Y field components at
various heights from the carpet surface. Effects due to finite electrode
widths only become manifest below around 0.25 pitches from the surface,
above which they quickly decay away. Right: Comparison of stable trajectory
heights in the idealized analytic calculation and full numerical simulation
of ions through the calculated E fields. At surface-to-ion distances
below around 0.25 pitches the higher order field terms cause ions
to crash into the carpet surface. This calculation is evaluated as
$m=136$ amu, $\mu_{0}=39.9$ cm$^{2}$V$^{-1}$s$^{-1}$, $E_{push}=20$ V cm$^{-1}$,
$V_{pp}=300$ V, $N=4$, corresponding approximately to the conditions
of barium ion transport in helium. \label{fig:Finite-electrode-width}}
\end{figure}

Inclusion of higher Fourier mode terms associated with the potential surface, which may be
associated with either finite-width electrodes or at sub-leading level with surface imperfections,  leads to another
important effect: the appearance of near-field effects at short distances from the carpet, which can destabilize the
micro-motion from circular orbits. These higher order terms will be
determined by the precise voltage structure on the carpet boundary,
and take the form in 3D space dictated by Laplace's equation:
\begin{equation}
\Phi(x,y,t)=\sum_{a}\Phi_{a}(x,y,t),\quad\quad\Phi_{a}(x,y,t)=\left[A_{a}(t)\sin\left(a\frac{2\pi}{Np}x\right)+B_{a}(t)\cos\left(a\frac{2\pi}{Np}x\right)\right]\exp\left(-\frac{2\pi}{Np}ay\right).
\end{equation}
We see from this expression how terms with $a>1$ decay away from
the surface more quickly than the leading $a=1$ sinusoidal term, which is
why the first order approximation is effective for describing dynamics
of the field at relatively large $y$. We can obtain an estimate of the
magnitude of the effects of these higher order terms by comparing
the idealized calculation of ion motion with a numerical simulation
in the full geometry, absent stochastic effects. An RF carpet geometry
with $160\,\mu$m pitch was implemented within the SIMION software
package and simulated in four-phase mode. Comparison of the leading-order
field configuration against the full finite-element simulated field
map from SIMION is shown at various surface-to-ion heights in Fig.~\ref{fig:Finite-electrode-width}
left. We observe that already by around 0.25 electrode pitches $p$
above the surface, the effects of the higher-order terms become vanishingly
small within the precision of the calculation. Simulating ion trajectories
through this numerically derived field configuration with stochastic
fluctuations disabled shows an excellent agreement with the analytic
expression at all surface-to-ion distances above 0.25$\times p$;
below this distance, ions begin to crash into the carpet due to higher
order perturbations to their micro-motion. Fig.~\ref{fig:Finite-electrode-width},
right, shows this comparison for a toy system with ion mass 136 amu
moving with mobility $\mu=39.9$ cm$^{2}$ V$^{-1}$ s$^{-1}$, $E_{push}$=20 V cm$^{-1}$
and $V_{pp}=300$~V RF voltage (the approximate conditions for barium
ion drift in helium buffer gas at 1 bar) as a function of pressure over an
$N=4$ carpet.\footnote{We note that to obtain this precision, SIMION had to be operated with
increased E field refine convergence of 0.0005 and enhanced tracking
quality T.Qual>20, above which precision the solution converged. Running
with previously recommended settings E field refine conference 0.005
and tracking quality T.Qual=0 lead to artificially strong performance
levitation of ions at short distances.} The blue band in Fig. \ref{fig:Finite-electrode-width} represents
the $\pm8.5\%$ RF voltage uncertainty described above, and has a
small effect. For any stable trajectory localized more the 0.25$\times p$
 from the surface, we conclude that finite electrode width
effects are negligible in determining the dynamics of phased RF carpets,
whereas below $y_{loss}=0.25\times p$ they generate near-field
effects that cause ions to crash into the surface.  The surface
 at $y_{loss}=0.25\times p$ can thus be considered
as being an effective surface-of-no-return for transverse ion transport
due to near-field effects.  The continued relevance of this surface in the presence of
stochastic effects will be an assumption of the thermodynamic model 
developed in later sections.

\section{Stochastic effects from Brownian motion and the equivalence of the
thermodynamic model \label{sec:Stochastic-effects-from}}

The calculations of Sec.~\ref{sec:Dynamics-of-ion} do not capture
the important effects of stochastic perturbations to the trajectory
caused by Brownian collisions with gas atoms or molecules. These effects
are especially critical for understanding the dynamics of systems
where the ion mass is large and the motion relatively viscous, as in
our systems of interest. These perturbations are the dominant cause of ion 
losses from the trapping region in high pressure gases.

There are multiple approaches to the problem of simulating buffer gas interactions
 with varying degrees of reliability. Within SIMION, two simulation methods are widely used.
An approximate ``statistical diffusion simulation'' (SDS)~\cite{appelhans2005simion,lai2008predictive}
method applies random impulses to the numerically evaluated
trajectory to simulate the integrated effect of many random interactions with the buffer gas. The SDS model
is expected to be valid in scenarios where ions travel in straight lines between collisions.
 Since inter-collision curvature is the hallmark of RF carpet operation that allows for
ion levitation, validity of the SDS model for our purposes appears questionable. Similar conclusions
have been reached by other authors, regarding use of the SDS model at radio frequencies and
high pressures~\cite{poteshin2020investigation,jurvcivcek2014design,pauly2014hydrodynamically}. A second
approach involves micro-physically simulating the trajectory collision-by-collision
with hard-sphere interactions~\cite{langridge2008simulation}. This
is more accurate but is far more computationally expensive.
Here we undertake simulations using the hard sphere model and show
that the effects of micro-physically simulated stochastic perturbations
can be well described using a novel thermodynamic and kinetic formalism
using the concept of thermalization into the pseudo-potential, allowing for much more rapid
evaluation of the viable parameter space.

\begin{figure}[t]
\begin{centering}
\includegraphics[width=0.535\columnwidth]{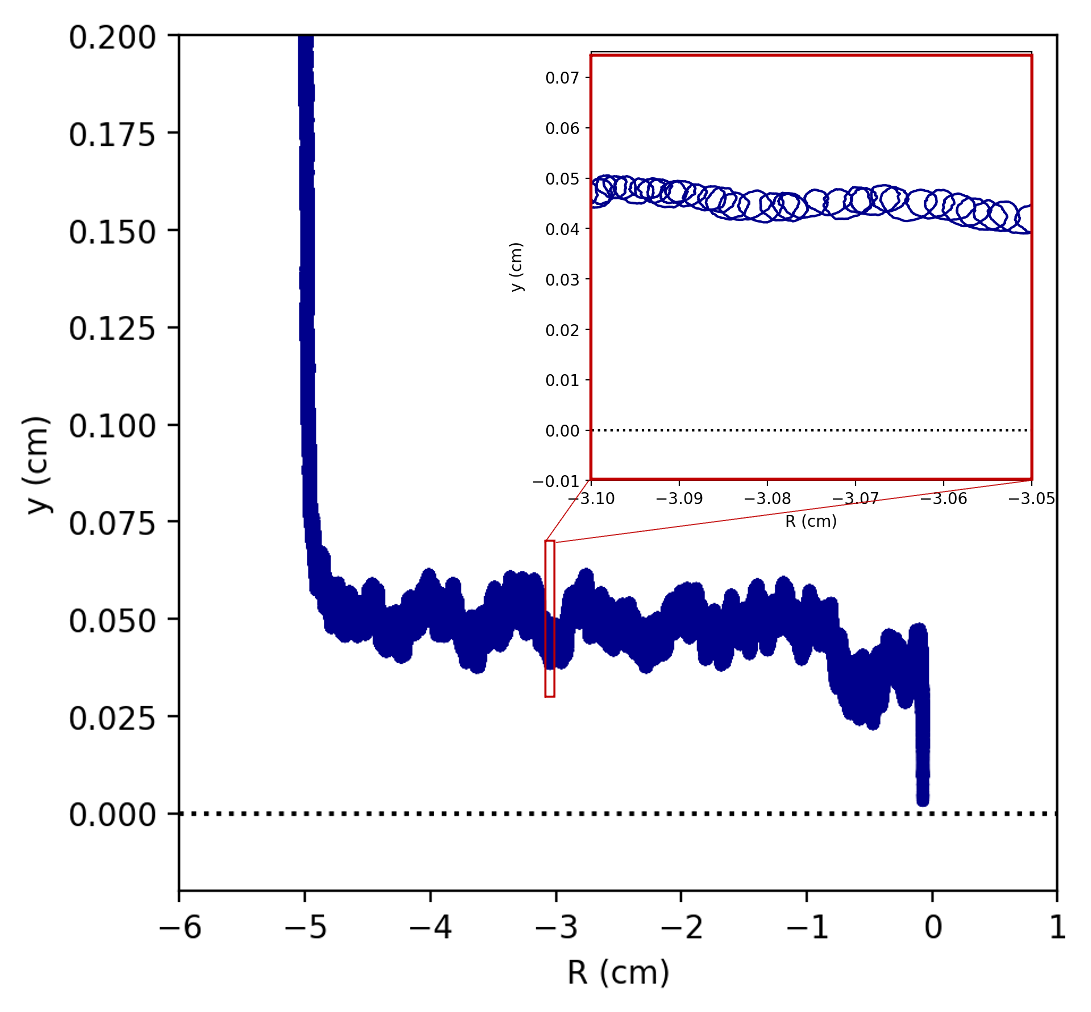}\includegraphics[width=0.46\columnwidth]{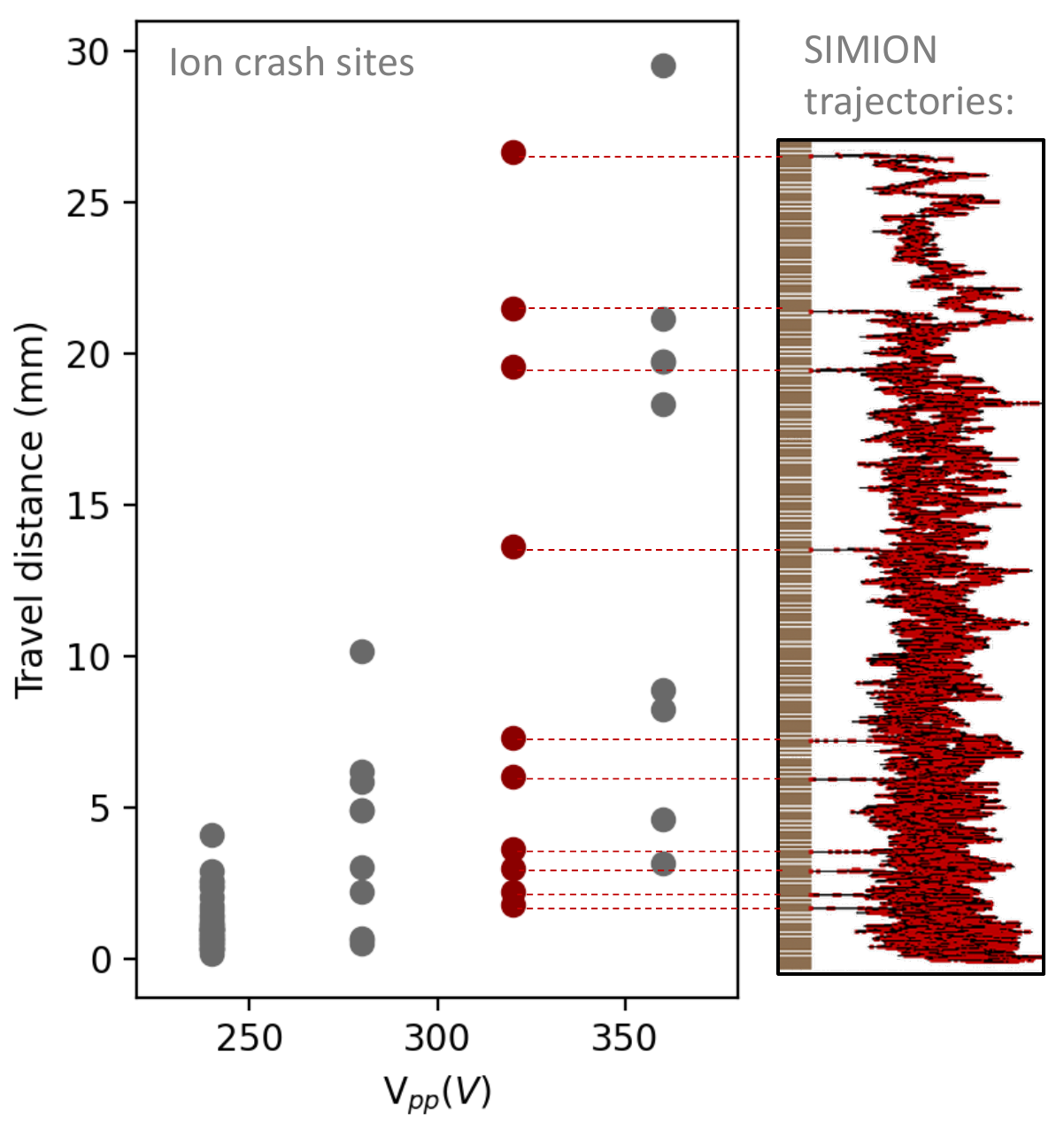}
\par\end{centering}
\caption{Left: Phased RF carpet simulation showing dynamics of micro-and macro-motion
for a successfully transported ion trajectory. Right: examples
of SIMION-simulated trajectories experiencing losses into the RF carpet
surface; Right: travel distance for ensembles of 10 ions at
various RF driving voltages which begin at $z=0$.  
The full SIMION trajectories for the 320~V data points are shown to the right for illustration.  In this sub-figure, $x$ runs vertically and $y$ horizontally, with ion crash sites mapped to the middle scatter plot.\label{fig:Left:-Phased-RF}}

\end{figure}

Figure \ref{fig:Left:-Phased-RF}, left shows a microphysical simulation
within the hard sphere model, using ion mass
$m=952$ amu (for reasons to be explained in Sec \ref{sec:Operating-Parameters-for}),
push field $E_{push}$=80 V\,cm$^{-1}$, buffer gas $M=136$ amu, collision
cross section $\sigma=6\times10^{-19}$ m$^{2}$, pressure $=1$ bar,
RF frequency $\Omega=13.56$ MHz, RF voltage $V_{pp}=700$ V, carpet
pitch $p=40\,\mu$m. This somewhat artificial simulation parameter
point was picked to show a regime where both the stochastic deviations
of the trajectory and the micro-motion are visible within a single
simulated trajectory.  In cases where one distance scale dwarfs the
other, illustrating the dynamics of the motion in a single image is
more challenging. The ion is placed high above the carpet and initially
travels directly downward with the push field. It then experiences
a transition to an RF carpet transport trajectory once the ion is near enough
the carpet to be significantly influenced by the RF forces.
We observe the characteristic spiral motion of the micro-trajectory
which causes levitation superposed on a random walk motion and a tendency
to move from left to right, in the direction of the traveling RF wave
phase. Eventually in this geometry the ion reaches the end of powered
part of the carpet and is driven down to the surface by the push field
near $R=0$.

To compare calculations with stochastic effects from the hard sphere
model to analytic models, we must have a full suite of parameters
for calculation of the pseudo-potential and the quantities that derive
from it. In most cases this is straightforward, though under hard
sphere dynamics the mobility of the ion is an emergent property rather
than an input parameter. Kinetic theory can be used to obtain a prediction
for this quantity in terms of the hard sphere model parameters,
\begin{equation}
\mu=\frac{3q}{16n}\sqrt{\frac{2\pi}{{\cal M}kT}}\frac{1}{\Omega_{M}}.
\end{equation}
Where ${\cal M}$ is the reduced mass, $n$ is the gas number density
and $\Omega_{M}$ is the momentum transfer cross section, which for
$m\gg M$ is equivalent to the geometrical cross
section for hard-sphere collisions. In terms of our input parameters:
\begin{equation}
{\cal M}=\frac{mM}{m+M},\quad\quad n=n_{0}\frac{P}{P_{0}},\quad\quad\Omega_{M}=\pi(2R_{Xe})^{2}.
\end{equation}
For the purposes of this demonstration we take $R_{Xe}$ to be the
Van Der Waals radius of xenon, $R_{Xe}=216\,\mathrm{pm}$ yielding
$\Omega_{M}=6\times10^{-19}$ m$^{2}$. We will take this prediction for
the mobility to complete the information needed to compare the microphysical
simulation with the analytic prediction. 

\begin{figure}
\begin{centering}
\includegraphics[width=0.49\columnwidth]{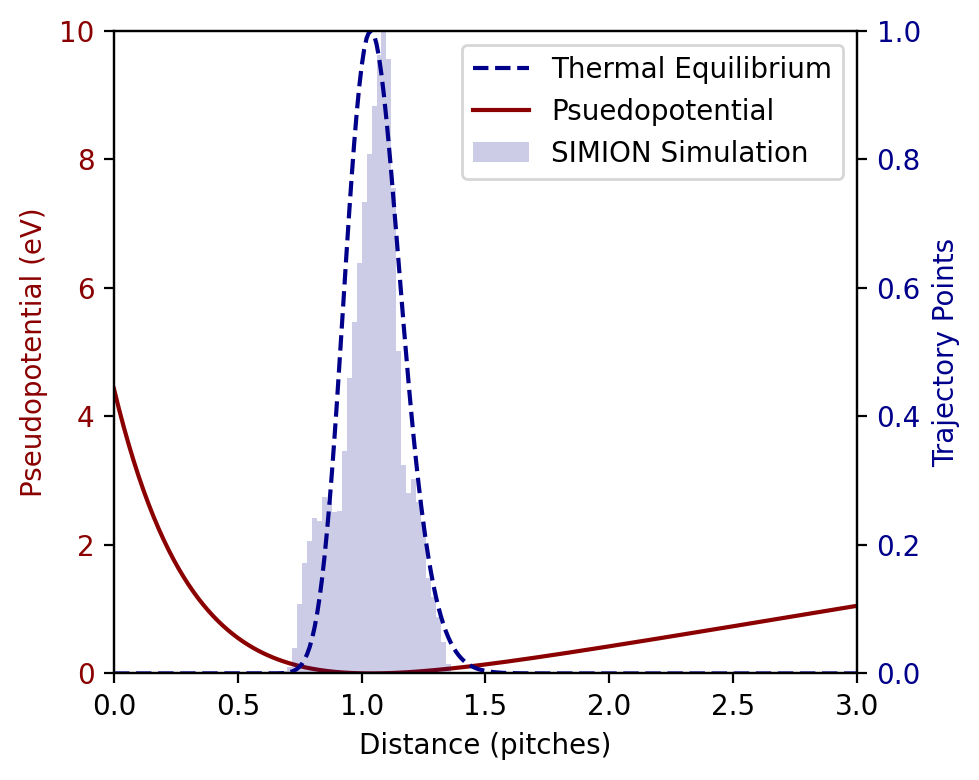}\includegraphics[width=0.49\columnwidth]{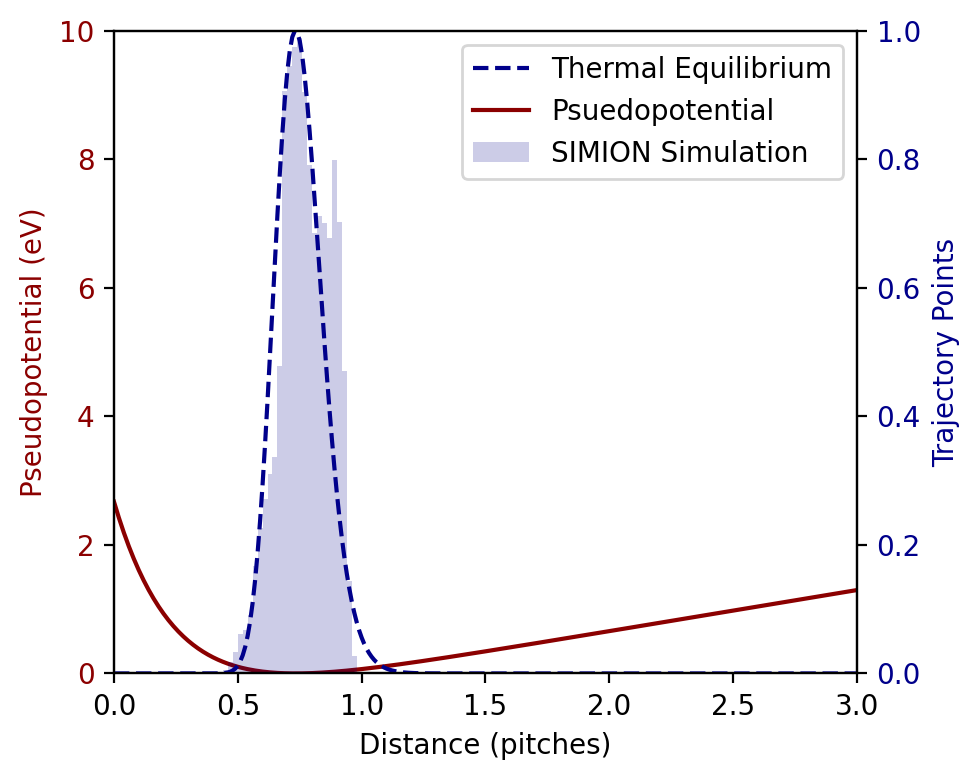}
\par\end{centering}
\caption{Demonstration of the validity of the thermalization conjecture for
(left) $N=4$,$V_{pp}=700V$ and (right) $N=3$,$V_{pp}=400V$. The
other operating parameters are $m=952$ amu, $M=136$ amu, $p=160\,\mu m$,
$P=1$ bar, $T=273$ K. \label{fig:Demonstration-of-the}}
\end{figure}

When stochastic effects are incorporated, the distribution of ion
heights above the carpet is no longer tightly bunched around the stable
macro-trajectory, but distributed around it by thermal motion. Since
ions in dense gases remain to a good approximation in thermal equilibrium,
we can consider the following conjecture: the ions undergoing collisions
with the Maxwellian distributed buffer gas should distribute with
a Boltzmann spectrum within the pseudo-potential, in the local rest frame of
the gas. This conjecture
is non-trivial, since the pseudo-potential treatment
technically only applies to ions following equilibrium trajectories,
and only in one dimension of 3D space in this system.  If the
thermalization conjecture is indeed satisfied, we would expect the
vertical height distribution of ions undergoing stochastic collisions
with buffer gas above the carpet to take the form:
\begin{equation}
\rho_{Therm}(y)=\frac{\exp\left[-qV(y)/k_{B}T\right]}{\int_{0}^{\infty}dy'\,\exp\left[-qV(y')/k_{B}T\right]}.
\end{equation}
Figure~\ref{fig:Demonstration-of-the} shows the comparison of the
height distribution projected from the thermalization conjecture with
the distribution predicted by detailed microphysical simulations for
two carpet configurations, one with three-phase RF driven at $V_{pp}=400$ V
and one with four-phase RF driven at $V_{pp}=700$ V. These two simulations
have parameters informed by the application of interest ($m=952$ amu, $M=136$ amu, $p=160\,\mu m$,
$P=1$ bar, $T=273$ K) but are shown 
for illustrative purposes only; precise calculations that properly include the pressure-dependent
mobility and ion mass will be given in later sections.
The thermalization conjecture is satisfied to a good degree of approximation
in both cases, despite different operational parameters. In simulation
scans over many conditions undertaken during this work we have observed
strong agreement with the thermalization conjecture in all cases.
The small deviations from the precise form of the distribution function
observed in Fig.~\ref{fig:Demonstration-of-the} are expected due to the finite
transport distances simulated.

The full simulation of ion trajectories, accounting for every collision
in dense gases, is an expensive and inefficient computational
prospect. The validity of the thermalization conjecture, however,
allows us to explore more rapidly the space of possibilities for
stable RF carpet operation at a given pressure and electric field
configuration. What we ideally wish to know is not only the ion height
distribution, but whether ions can travel for large distances along
the carpet surface without losses. 

First, we note that it is obvious that when there is no stable trajectory,
or $\langle y\rangle<0$, all ions will be quickly lost onto the carpet surface.
Rearranging this condition we can find a minimal value of $V_{pp}$
for transport to be plausible,
\begin{equation}
V_{pp}\geq\sqrt{\frac{m\left(D^{2}+\tilde{\Omega}^{2}\right)E_{push}}{2q}\left(\frac{Np}{\pi}\right)^{3}}.
\end{equation}
Above this driving RF voltage, stable transport is possible because
the pseudo-potential well is deepest at some position above the carpet
rather than below it. However, there will still be losses due to stochastic
effects at finite buffer gas densities: no matter how strong the
effective repulsive force, the equilibrium distribution will
have some small tail that runs below the surface height. These losses
are manifest as ion crashes after random distances, with the mean
distance depending on the applied RF voltage. Fig.~\ref{fig:Left:-Phased-RF},
right shows an example set of simulations at various RF voltages using the conditions 
described above with the random crashes at various distances illustrated. 

Ion losses are substantial when the equilibrium probability distribution
of ion heights has a substantial contribution at or below the surface
beyond which ions are lost, established in Sec.~\ref{sec:Finite-electrode-width}
to be around $y_{Loss}=0.25\times p$. Thus the condition for long-lived
ion levitation can be expressed as
\begin{equation}
\phi=\frac{\int_{-\infty}^{y_{Loss}}dy\,\rho_{Therm}(y)}{\int_{-\infty}^{\infty}dy'\,\rho_{Therm}(y')}\ll1.
\end{equation}
While qualitatively interesting for understanding the conditions for stability,
this argument is not very useful in practice, since it is unclear how we
should interpret $\ll$ quantitatively. The ion loss rate depends
not only on the thermal distribution, but also on how fast
ions explore that distribution. To obtain an estimate for the theoretical
ion loss rate for the potential we must consider ion kinetics as well
as thermodynamics.

For simplicity of exposition in the following we assume ions to be
lost at the surface $y=0$, though in practice we will set this value
equal to $y_{Loss}$ for predicting ion loss rates in the calculations
that follow. At sufficiently high pressures we can consider that the
trapped ions are undergoing constant collisions with the buffer gas,
maintaining thermal equilibrium. To a good approximation each scattering
interaction causes the ion to obtain a random velocity in a random
direction, sampled from the Maxwell-Boltzmann distribution. Consider
now the motion of an ion between two collisions separated by mean
collision time $T$. To hit the wall following such a collision an
ion at position $y$ must have a velocity of at least $v_{y}=y/T$
in the $y$ direction. Thus between these two collisions the probability
for ion loss is

\begin{figure}[t]
\begin{centering}
\includegraphics[width=0.49\columnwidth]{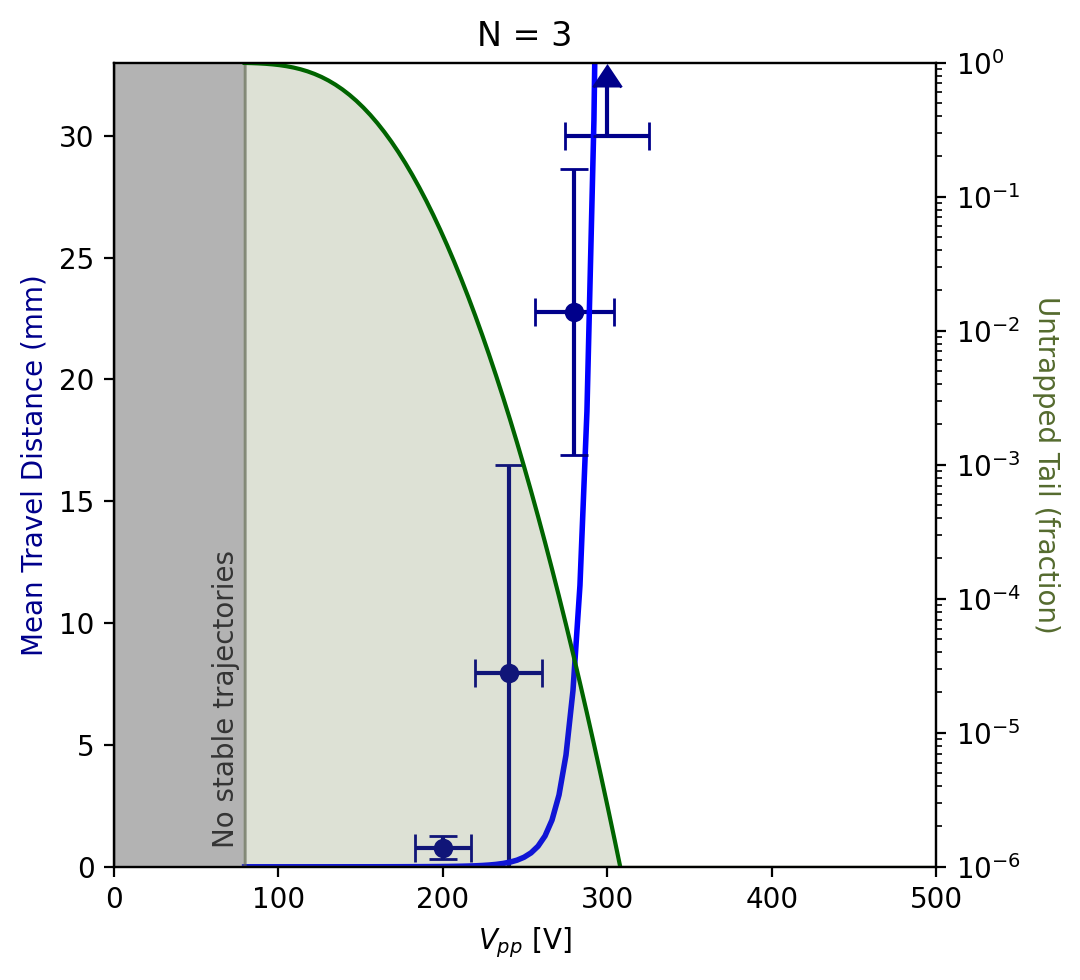}\includegraphics[width=0.49\columnwidth]{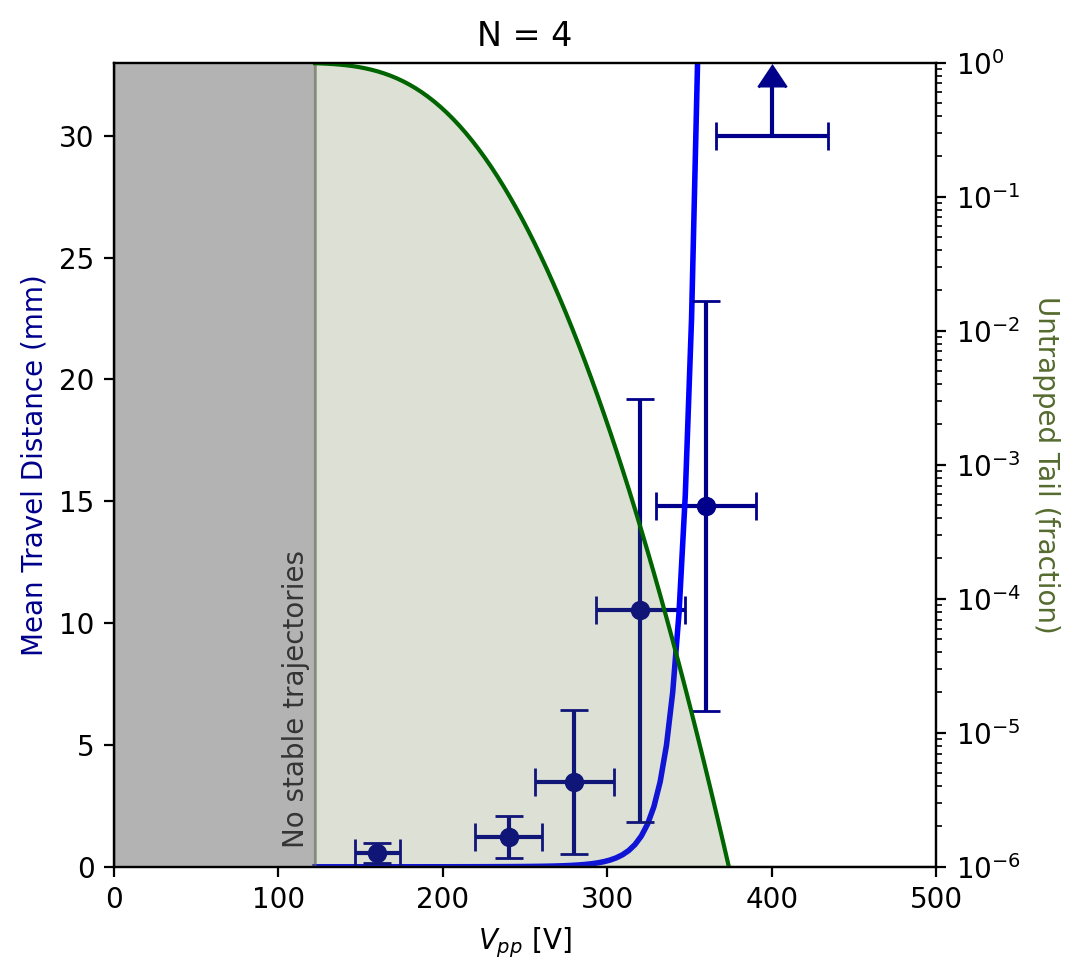}
\par\end{centering}
\caption{Comparison of simulated and theoretical ion survival distances for
N=3 and N=4 phased arrays at 160 $\mu m$ pitch, $m=952$ amu, $M=136$
amu, $p=160\,\mu m$, $P=1$ bar, $T=273$ K. The grey shaded area represents
the region where $\langle y\rangle$ < 0 so no stable trajectories exist. The
green curve shows the integral of the negative tail of the thermal
distribution below $y_{Loss}=0.25\times p$, to be read using the
right logarithmic axis. The blue line shows the predicted mean ion survival distance,
assuming validity of the thermodynamic arguments
of this paper. The blue data points show the results of a detailed,
collision-by-collision simulation in the SIMION software package.
The error bars show mean and standard deviation over a sample of an
ensemble of simulated ions. Strong qualitative agreement between the
theoretical model and the microscopic simulation is observed, in particular
in specifying the location of the sharp transition from lossy to relatively
lossless transport. \label{fig:Comparison-of-simulated}}
\end{figure}
\begin{equation}
{\cal P}(T)=\int_{-\infty}^{0}dv_{y}f(v)\int_{0}^{v_{y}T}dy\,\rho(y).
\end{equation}
If the collision time is short, as will always be the case in high
density buffer gases, we can approximate that $\rho(y)=\rho(0)$ everywhere
the above integral has support. Thus the probability for an ion loss
between these two collisions can be evaluated as
\begin{equation}
{\cal P}(T)=\langle v_{y}^{-}\rangle T\rho(0).
\end{equation}
In this expression, $\langle v_{y}^{-}\rangle$ is the mean negative
velocity in the $y$ direction, which can be evaluated from Maxwell-Boltzmann statistics,
\begin{equation}
\langle v_{-}\rangle=\int_{-\infty}^{0}dv\,v\sqrt{\frac{m}{2\pi kT}}\exp\left(-\frac{mv^{2}}{2kT}\right)=\sqrt{\frac{kT}{2\pi m}}.
\end{equation}
Finally we can calculate the expected loss rate per unit time $\Gamma_{t}$
and per unit distance along the carpet $\Gamma_{d}$ respectively:
\begin{equation}
\Gamma_{t}=\frac{1}{T}{\cal P}(T)=\sqrt{\frac{kT}{2\pi m}}\rho(0),\quad\quad\quad\Gamma_{d}=\frac{1}{\langle v_x\rangle}\Gamma_{t}.
\end{equation}
Thus we obtain statistical predictions of ion loss rates in distance
and time, based on the previously demonstrated thermalization conjecture
augmented with kinetic arguments. Figure~\ref{fig:Comparison-of-simulated}
shows the comparison of simulated ion loss rates from a complete collision-by-collision
SIMION simulation to the thermodynamic prediction, for similar operating
parameters to atmospheric pressure xenon gas operation. 
Quantitative agreement is observed. In particular, while the ion
survival distances in the transition region are highly variable, the
sharp transition from stable to unstable transport around a certain
well-predicted critical voltage is reproduced in both the analytic
model and simulation.

\section{Comparative analysis of $N-$phased array performance characteristics
\label{sec:Comparative-analysis-of}}

\begin{figure}
\begin{centering}
\includegraphics[height=0.35\textheight]{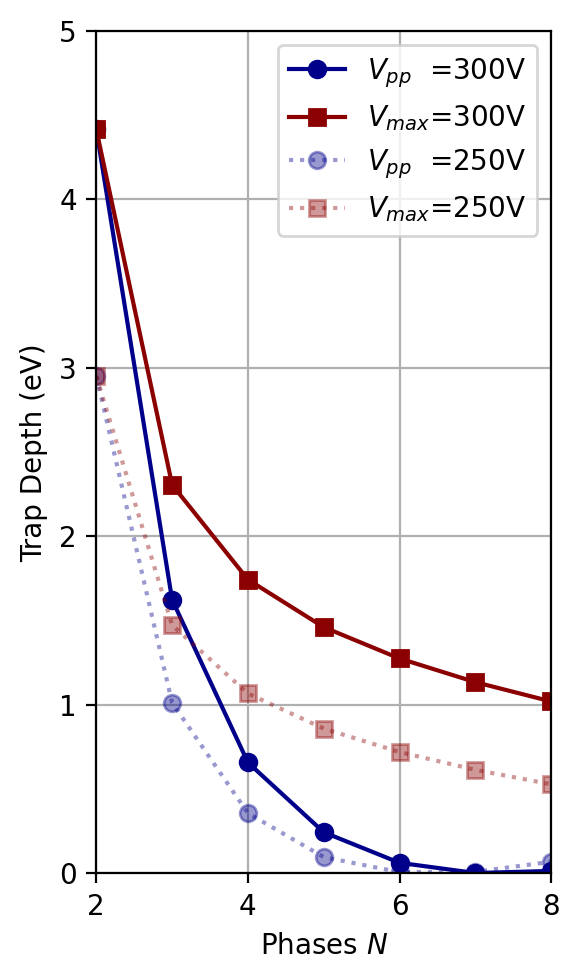}\includegraphics[height=0.35\textheight]{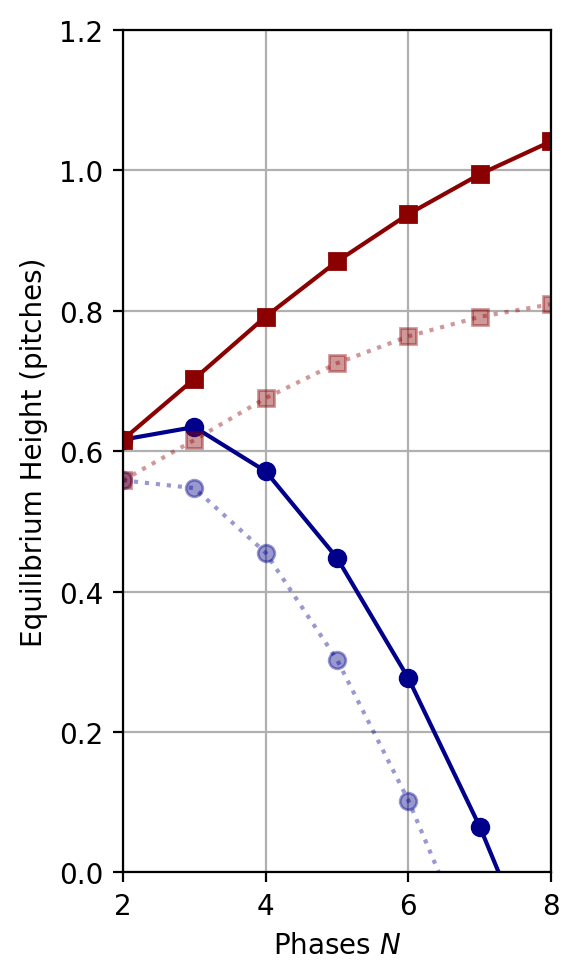}\includegraphics[height=0.35\textheight]{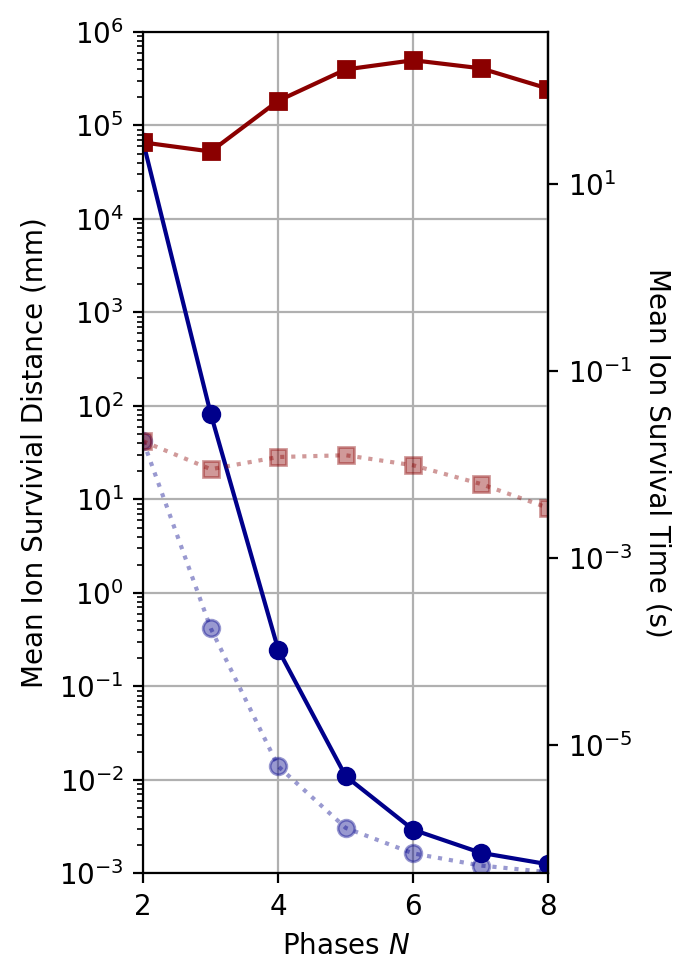}
\par\end{centering}
\caption{Dependence of RF carpet stability parameters on phasing N for a 160~$\mu$m carpet. The left
plot shows trap depth in eV, center shows equilibrium ion transport
height (the minimum of the pseudo-potential) and the right plot shows
the mean survival distance and lifetime, calculated according to the
thermodynamic protocol outlined in this paper. The operating parameters 
are $m=952$ amu, $M=136$ amu, $p=160\,\mu m$,
$P=1$ bar, $T=273$ K. \label{fig:NDependence}}
\end{figure}

In the previous sections we have proven that the important properties
of phased RF carpet ion transport, including stochastic effects from
ion-gas interactions leading to ion losses into the carpet, can be
modeled analytically and without computationally expensive simulation
ensembles.

In this section we demonstrate why phase orders $N\geq3$ may be considered
compelling relative to the simpler $N=2$ system, already discussed
by Schwarz \cite{schwarz2011rf}. As shown in the left two panels of Fig.~\ref{fig:NDependence} (blue
lines), for the same fixed RF driving voltage $V_{pp}$, trap depth
and equilibrium ion height both will generally fall with $N$. This
naturally implies a reduced mean ion survival distance, as shown in
Fig.~\ref{fig:NDependence}, right (blue lines), which would seem to
imply $N=2$ as an optimum operating phase structure. Since the mean
transport velocity Eq.~\ref{eq:TransVel-1-1} does not depend on $N$
or $V_{pp}$ we label this figure with both mean survival time and
survival distance axes, though with the caveat that the $N=2$ data
point should technically only be interpreted according to the mean
survival time metric, since in this case Eq.~\ref{eq:TransVel-1-1} is inapplicable 
given no tendency to drive the ion in one direction or the
other. 

The highest voltage that can be applied to an RF carpet is limited
by breakdown, either through the buffer gas or through the carpet
insulating material. In an $N$-phased system, the maximal voltage
between any adjacent pair of electrodes scales as:
\begin{equation}
V_{max}=V_{pp}\sin\left(\frac{\pi}{N}\right).
\end{equation}
which is less than $V_{pp}$ for a $N\geq2$ phased system (see Fig.~\ref{fig:VMax}).
This relationship can be inverted to provide a maximal allowed drive
voltage for an $N$-phase carpet limited by material dielectric strengths.
The higher driving voltages that can be applied to $N\geq2$ phased
arrays then compensate for some of the performance degradation introduced
by reducing the levitating field gradients. This is shown in Fig.~\ref{fig:NDependence}
(red lines). While the trap depth continues to drop with increasing
$N$ for fixed $V_{max}$, the equilibrium ion height above the carpet
increases; this countervailing effect leads to a non-trivial dependence
of mean survival distance on $N$, as shown in Fig.~\ref{fig:NDependence},
right. For an operating voltage of $300$ V and a pitch of 160 $\mu$m,
the optimal array phasing appears closer to $N=6$. Practical issues
such as arranging for supply of $N$ distinct RF phases to one small-pitch
carpet structure or power dissipation considerations may favor smaller $N$ than this optimal
choice, but from the point of view of ion transport dynamics alone,
intermediate values for the carpet phasing are preferable, contrasting
to some degree our conventional intuition.

\section{Application of phased RF carpets for neutrinoless double beta decay\label{sec:GeneralConsideration}}

\begin{figure}
\begin{centering}
\includegraphics[width=0.85\columnwidth]{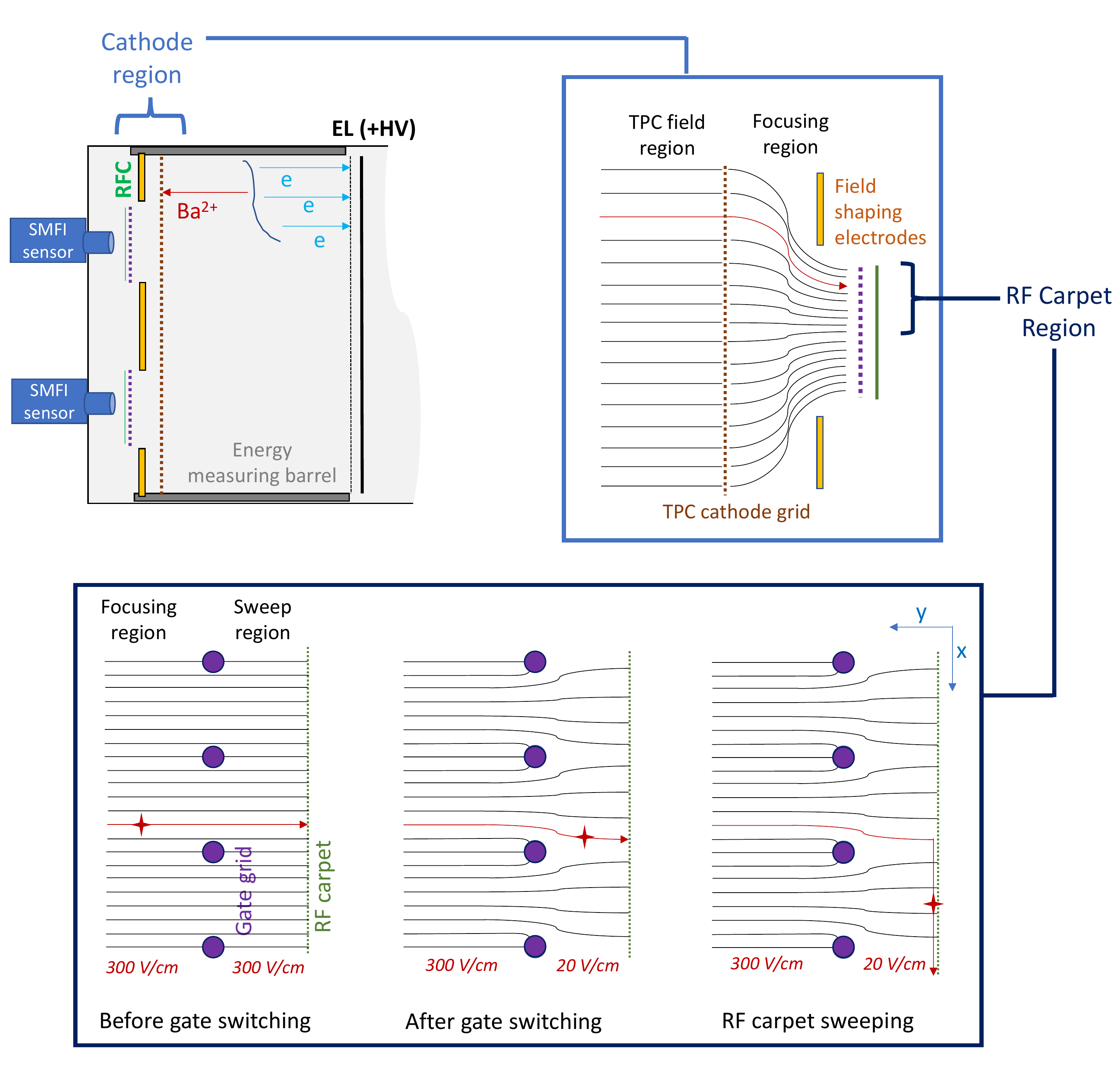}
\par\end{centering}
\caption{Sketch of a possible RF carpet implementation for a xenon gas double beta decay search.  Thin black lines show a cartoon of the electric field lines (though concentration and expansion factors of the fields will likely be much larger than indicated here). The top left figure shows an outline of the cathode and its relation to other parts of the detector.  Electrons drift rapidly to the anode where they produce electroluminescence light in \cal{O}(1)~ms after the event, whereas ions drift to the cathode in \cal{O}(1) second where they are captured in the RF carpet sweep region by a switching gate grid.  \label{fig:NEXTSketch}}
\end{figure}

For a ton-scale high pressure xenon gas detector operating at 10 bar, similar to
the system\footnote{Ref.~\cite{adams2020sensitivity} proposes 15 bar operation. The same detector system could, however, be operated at 10 bar, which we have chosen as the benchmark pressure for this work to match the operating pressure of existing NEXT detector phases.} described in Ref.~\cite{adams2020sensitivity}, barium
ions will naturally drift in the applied electric field to the cathode
plane, which is likely to be of order 2 m in diameter. RF carpets
have been demonstrated to efficiently transport ions over similar distance scales~\cite{savard2011large}. However, the highest operating
pressure investigated to date has been 300 mbar.  RF carpets for use
at higher gas pressures would require much smaller pitches and higher applied RF voltages than
have been previously demonstrated. 
Nevertheless, several arguments suggest that such an application is not
entirely unfeasible: 
\begin{enumerate}
\item High pressure xenon gas has a breakdown voltage dramatically in excess
of helium, so larger RF operating voltages can be employed. This enhancement
was tested experimentally in Ref.~\cite{woodruff2020radio}, which ultimately
proved that dielectric materials such as kapton will likely break
down before the gas in the detector at spacings relevant for RF carpet operation. 
\item Molecular ion formation around dications in xenon gas is expected
to enhance their mass for the purposes of transport at high pressure.
This physics was elaborated and a detailed theoretical presented
in Ref.~\cite{bainglass2018mobility}.  This effect is especially strong for
doubly charged ions, leading to an important stabilization effect in RF transport.
\item Technologies to produce large RF carpets at somewhat smaller pitches
than the present state of the art appear viable, employing nano-fabrication
methods.  Scaling 
to large radii is a manufacturing challenge, but not an obviously insurmountable one.
\end{enumerate}
While detailed design of a barium tagging phase of of xenon gas double
beta decay program remains both speculative and outside the scope of the present work, it is possible to consider
a set of plausible requirements for an RF carpet array used as part
of a barium tagging sub-system.  A sketch of one possible implementation is shown in Fig.~\ref{fig:NEXTSketch}.

A single RF carpet of 2~m radius is not required for a barium tagging cathode, as long as ions arriving anywhere
on the cathode surface are captured and transported to a sensor.  Several practical requirements on the RF carpet system are
eased both by modularizing the barium sensing system into sub-units and employing electrostatic focussing to
the extent possible.   In the scheme of Fig.~\ref{fig:NEXTSketch}, ions first drift to the cathode where they are guided by
static fields to an array of RF several carpets, each concentrating ions to one
single molecule fluorescent imaging ion sensor. A full coverage of the $\sim$2~m diameter cathode with seven RF carpet sub-units, with electrostatic focussing by a factor of four in field strength implies the need for seven 30~cm diameter RF carpets, which would demand efficient ion transport for at least 15~cm of distance.   Stronger electrostatic focussing may in principle be viable before the TPC operation becomes unstable, though with some active gas volume sacrificed for the focussing volume. 

The required drift field of a time projection chamber is larger than the small
$E_{push}\sim 20$ V cm$^{-1}$ required for RF carpet operation to be efficient, 
even without electrostatic focussing.  Thus the ions must necessarily travel
from a high-field region of the TPC into a low-field region of the RF carpet volume. 
 In an electrostatic system this
would imply losses of ions on whatever gate structure separates these regions. 
In an electrodynamic mode, however, ion losses through this transition
can be averted by timed gating.  In an electroluminescent
 time projection chamber the projected arrival time of an ion at the carpet plane can be
straightforwardly established online by comparing the time difference between primary (S1)
and secondary (S2) scintillation signals generated by the much faster drifting electrons.  These 
signals are both acquired by the energy detection system, which
is a mandatory component of any neutrinoless double beta decay time projection chamber experiment.
 Upon arrival of the ion within the RF carpet sweep region, the local electric field  
can be switched from the high value required for efficient drift to the lower value required for
a stable push-field.  This switch can be applied at trigger level, without detailed online event reconstructions, for all events within a broad energy energy region of interest based on uncalibrated S2 pulses. Neither the voltage difference ($\pm$ a few hundred volts) 
nor the required speed  ($\sim$ 1~ms) or timing precision ($\sim$ 0.1~ms) of this switch appear to be prohibitive, and it does not interfere with the electric field in the detector drift region, which remains stable, although temporarily insensitive to barium ions.

There are many conceivable arrangements of the RF carpet system geometry, with varying details of coverage, shape, and segmentation. One factor that is likely to strongly influence the choice of geometry is power consumption, which must be managed with care.  For carpets with the fine pitches that will be considered for our application of interest, resistive losses are the dominant 
source of power consumption over dielectric ones.  The power lost to resistive heating depends on the radius as well as the pitch and connectivity of RF electrodes.  The power dissipated in a single, fully connected 2 m diameter, 10 $\mu$m pitch carpet operated at 1000~V and 1~MHz is likely to be unviably large (potentially at the 1~MW 
level, though with significant dependencies on choice of both insulator and conductor materials). Dividing the carpet into smaller sub-units and reducing their surface coverage can reduce this power consumption considerably. Resistive heating of a fully connected RF carpet scales with approximately the forth power of the radius, with both the total capacitance and total length of conductor factoring into the heat load.   The modularized scheme as described therefore implies a significantly reduced power consumption relative to a single fully carpeted cathode surface, estimated to be of order 4~kW for the set of seven fully connected, 30~cm diameter RF carpets.  Azimuthally segmenting the ring electrodes into six segments per carpet could further reduce the resistive heating for the same coverage to around 100~W by reducing the lengths of each conductor that must be charged and discharged. While it is beyond the scope of this work to attempt a detailed optimization of system geometry, or to present detailed power-draw calculations,  it is recognized that power consumption is likely to be a relevant consideration for a final design in addition to those presented here, likely managed primarily through system geometry optimization.  System considerations may also limit the deployable number of RF phases.

Finally we note that, since establishing coincidence between a detected double beta decay event and the collected barium
ion is mandatory for barium tagging, transport speed along the carpet surface is also a relevant  performance parameter.
 Given the low rate of events
expected with energy near $Q_{bb}$, a time delay from landing to arriving
at a sensor of 10 seconds can be easily accommodated. All efficient
transport scenarios we explore will meet this requirement comfortably, so
ion delivery speed is not a considerable design driver.

\section{System parameters for RF carpets in high pressure xenon gas \label{sec:Operating-Parameters-for}}

Following the previous sections of general discussion we now turn
attention to the RF carpet operating parameters that may be appropriate
for barium ion transport in high pressure xenon gas. These inputs
to this calculation can be divided into two categories. First, the
input parameters associated with the medium or ion, which are fixed
by nature and we cannot change; second, the input parameters we may
adjust by choice of technology or operating conditions. 

\subsection{Gas and ion parameters}

The following parameters are considered intrinsic properties of the
system, and will not be adjusted during this study. 

\medskip{}

\textbf{Gas composition / atomic mass $m$:} The gas in a high pressure
xenon gas neutrinoless double beta decay experiment can be assumed
to be predominantly $^{136}$Xe with an atomic mass of 136. Possible
addition of minority components to reduce diffusion or absorb neutrons,
such as $^{4}$He~\cite{XePa,Felkai:2017oeq,Fernandes:2019zuz,McDonald:2019fhy} or
$^{3}$He~\cite{rogers2020mitigation} respectively may also be present.
Unlike electrons, ions are always thermalized at the drift fields
of interest, and so a minority component of a light noble gas will
not impact either the instantaneous energy spectrum or bulk mobility
of drifting ions, and we can assume $m=$136 amu for the mass of the
buffer gas with a reasonable expectation of accuracy for all these
cases. Incorporation of a molecular gas such as CH$_{4}$~\cite{henriques2019electroluminescence}
or TMA~\cite{gonzalez2015accurate,cebrian2013micromegas,trindade2017experimental,trindade2018study}
and others such as CO$_2$ ~\cite{henriques2017secondary} and CH$_4$~\cite{azevedo2016homeopathic} have also been proposed, though this requires more involved filtering
schemes to purify the working gas mixture to the part-per-billion
levels of purity in oxygen and water required for TPC operation. Incorporation
of such a molecular species would lead to non-trivial chemistry forming
molecular ions with the drifting barium cluster, and so we do not
consider it here.

\begin{figure}
\begin{centering}
\includegraphics[width=0.99\columnwidth]{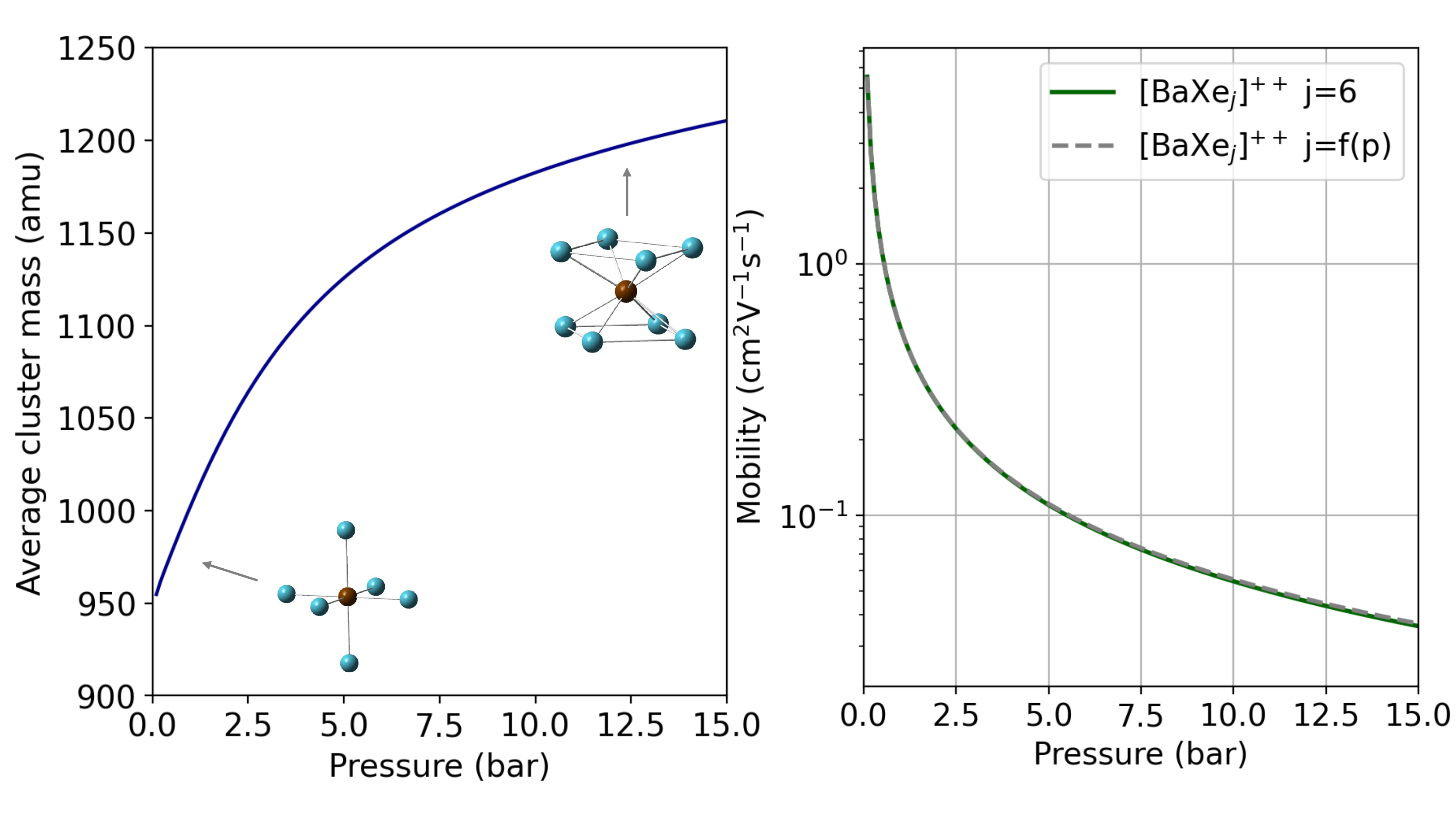}
\par\end{centering}
\caption{Left: mean cluster mass vs. pressure using methods of Ref \cite{bainglass2018mobility}.
Right: mobility vs. pressure from Ref \cite{bainglass2018mobility}.
The pressure dependencies of these parameters have important implications
for the pressure dependence of RF carpet operation in the pressure
range 1-10 bar. \label{fig:Left:-mean-cluster}}
\end{figure}

\medskip{}

\textbf{Ion charge} $q$: Barium ions produced in any double beta
decay event are created in a very highly charged state as emerging
electrons disrupt the decaying atom. Within picoseconds these highly
charged initial ions have captured electrons from surrounding neutral
xenon atoms until this capture process becomes energetically disfavored,
a process which terminates at Ba$^{2+}$. In high density environments
such as in liquid or solid xenon, recombination of thermalized charge
near the ion leads to further neutralization, leading to a distribution
of barium charge states. In xenon gas at densities up to at least
50 bar, recombination is expected to be minimal, as illustrated in
Refs.~\cite{Bolotnikov} and~\cite{novella2018measurement}, and so
the ions can be assumed to be in the pure 2+ charge state.

\medskip{}

\textbf{Ion mass} $M:$ The mass of the bare daughter ion $^{136}$Ba
is 136 amu. This ion has a high charge density and xenon atoms are
easily polarizable, and thus the ion quickly becomes dressed with
a shell of neutral xenon atoms. Calculations of molecular ion formation
in the Ba$^{+}$/Xe system using density functional theory~\cite{bainglass2018mobility}
agree very well with experimental data which show a strongly pressure
dependent reduced mobility, the hallmark of molecular ion formation~\cite{Cesar2014}. In the Ba$^{2+}$/Xe system, similar calculations
predict formation of significantly larger clusters, with a mean Xenon
binding number of 6--9 depending on gas pressure. The binding energies
are 3--4 eV, far above thermal energies, suggesting these clusters
should be thermally stable as they are driven by the RF carpet. The
lightest cluster predicted has M=952 amu, and this was used as a reference
parameter point in the earlier sections in this paper for this reason.
In this calculations that follow we use the predicted pressure-dependent
mean cluster mass extracted from the simulation results of Ref.~\cite{bainglass2018mobility},
shown in Fig.~\ref{fig:Left:-mean-cluster}, left. In general, both
the trap depths and mean ion travel heights improve for larger values
of $M$ so the expected molecular ion formation can be considered
as a beneficial effect for RF carpet operation in high pressure xenon.

\medskip{}

\textbf{Ion mobility} $\mu$: In Ref.~\cite{bainglass2018mobility},
potential energy surfaces evaluated with density functional theory
were used to calculate the momentum transfer cross section between
neutral xenon atoms and the expected ion clusters in high pressure
xenon gas. These cross sections were then used to obtain the ion
mobility. In the case of the Ba$^{+}$/Xe system, the relative mass
of the average cluster changes dramatically with pressure in the 0.1--10
bar range, since the mass of bare barium and {[}BaXe{]}$^{+}$ are
very different, and this substantially influences the reduced mobility
as a function of pressure. However, for the Ba$^{2+}$/Xe system the
relative mass of the cluster is less dependent on pressure and hence
the reduced mobility varies less drastically. Fig.~\ref{fig:Left:-mean-cluster},
right shows the predicted mobility, either assuming a fixed cluster
size of 6 (dashed grey line) or accounting for the predicted pressure-dependent
scaling of cluster size (green line). The difference between these
predictions is small, with absolute mobility scaling with a fairly
constant reduced mobility $\mu_{0}$ as $\mu=\mu_{0}\frac{p_{0}}{p}$.
This pressure dependence of absolute mobility leads to increased damping
at high pressures, however, and can be understood as one of the fundamental
reasons why RF carpets become more challenging to use at high operating
pressures.

\subsection{RF carpet parameters and their constraints}

The following parameters are adjustable based on design choices, and
our goal in this work is to establish the values  that would allow for stable transport in 
barium tagging applications:

\medskip{}

\textbf{Pitch }$p$: The RF carpet pitch defines the distance between
two electrode centers. The smaller the pitch, the stronger the levitating
effect of the RF carpet. Present large RF carpet manufacturing techniques
aiming for fine pitches have used a single layer PCB etch on Kapton
substrate with pitches of around 160 $\mu$m~\cite{arai201456}.
State-of-the-art PCB manufacture is capable of still smaller
pitches, down to 25 $\mu$m trace and gap width~\cite{Averatek}, yielding 50 $\mu$m pitch.
Manufacturing methods for RF structures used in ion traps for quantum
information processing have, however, achieved far finer feature sizes
than 50 $\mu$m~\cite{sterling2014fabrication,blain2021hybrid,seidelin2006microfabricated,brown2021materials,niedermayr2014cryogenic}
and manufacture by photo-etch on sapphire or by CMOS production both
appear possible for achieving feature sizes as small as 5$\mu$m~\cite{mehta2014ion}. RF carpets
of this scale have been proposed, but not yet demonstrated, by others~\cite{poteshin2020investigation}.  
Ongoing R\&D within the NEXT collaboration
aims to produce an RF carpet by single-layer photo-etch on sapphire
of pitch of 15$\mu m$ for experimental studies of ion transport in
xenon gas.

\medskip{}

\begin{figure}
\begin{centering}
\includegraphics[width=0.5\columnwidth]{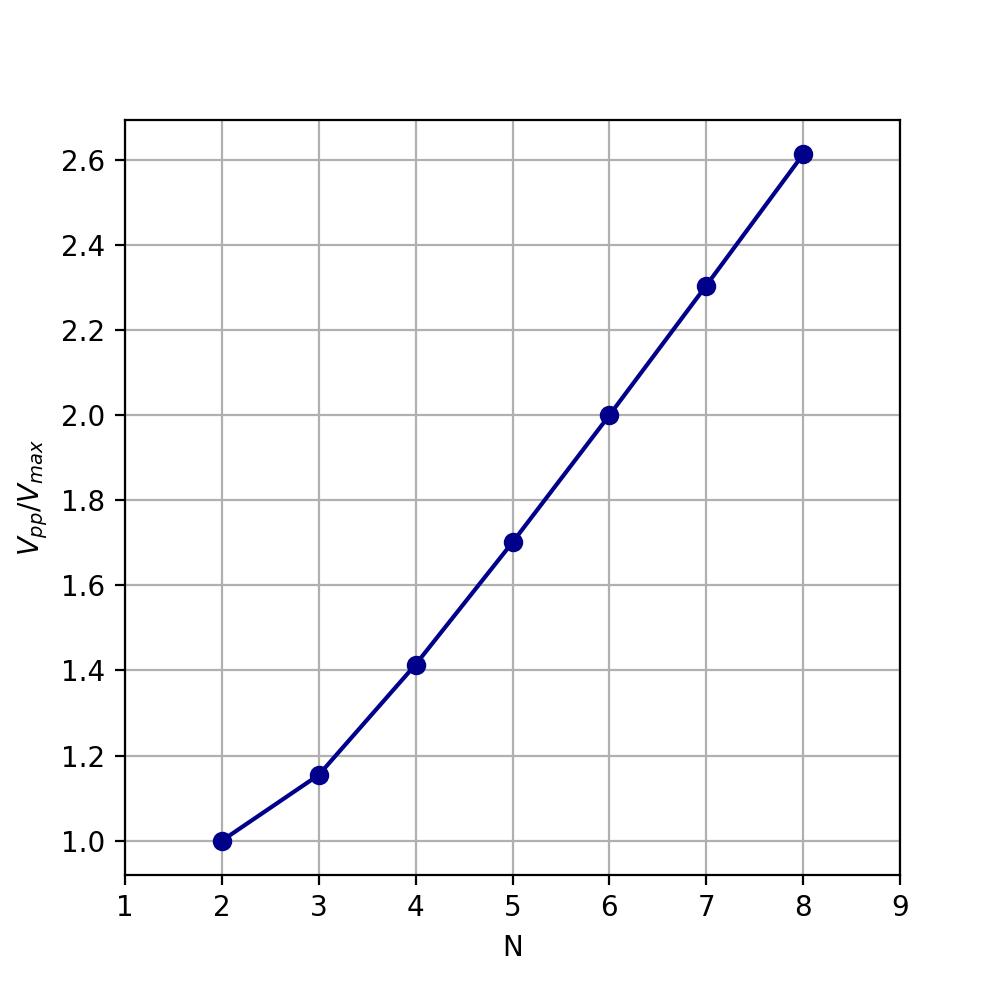}
\par\end{centering}

\caption{Dependence of maximal electrode-to-electrode voltage $V_{max}$ with
carpet phasing $N$ for a fixed peak-to-peak voltage $V_{pp}$\label{fig:VMax}}

\end{figure}

\textbf{Carpet phasing $N$: }In principle any phasing is possible, though
the complexity of supplying the various RF voltages to independently
biased groups of traces becomes more difficult as the phasing gets
larger. $N=2$ phasing is the simplest option, though this does not
allow for transverse motion without also applying either a DC sweep
field or a surfing wave. A DC sweep field is not ideal for
use in a NEXT detector since it would require one carpet edge to be
highly biased relative to the ion collection point, under which conditions it would
be challenging to maintain a sufficiently small push field across the array without a rather complex gating geometry. 
The study of Sec.~\ref{sec:Comparative-analysis-of} showed $N=3$ gave particularly
poor performance. We thus pick $N=4$ as our benchmark. Calculations for $N=5$ and $N=6$
show minimal improvement in the transport efficiency for small
pitches, certainly insufficient to justify the additional complexity
associated with a 5 or 6 phase RF structure.

\medskip{}

\textbf{RF peak-to-peak voltage $V_{pp}$:} The RF peak-to-peak voltage
is limited by the breakdown voltage for discharges over the carpet
surface or through the insulating material(s), at some threshold voltage\textbf{
$V_{max}$}, as described in Sec.~\ref{sec:Comparative-analysis-of}
and shown in Fig.~\ref{fig:VMax}.  For the $N=4$ carpets that 
will be used as example cases in the following sections, the ratio of V$_{max}$ to V$_{pp}$ is around 1.4. 
Generally speaking, breakdown through the buffer
gas has limited operating voltage of all previous RF carpets. Typically
in helium gas it is possible to reach applied voltages around \textbf{$V_{pp}$}=150 V.
Xenon gas has a far higher breakdown strength than helium gas, and
elevated pressures increase this breakdown voltage still further.
Past measurements of the breakdown of thin Kapton layers at RF frequencies
in xenon gas suggest that in high pressure xenon gas experiments,
breakdown of insulator substrate will be limiting before breakdown
through the gas for kapton PCBs operated at gas pressures above 1
bar. Data from Ref.~\cite{woodruff2020radio} showed breakdown occurring
through kapton films at approximately $V_{max}=500$ V for 130 $\mu$m
insulator distance, though it is likely that in a practical RF carpet this value may
be somewhat higher due to the higher quality of the electrode surfaces.  Several commercially
available kapton films advertise a dielectric strength as high as 8~kV for
a 25 $\mu$m film thickness, suggesting that with suitably well-prepared electrodes,
much larger voltages may be supportable on kapton substrates~\cite{KaptonHV}.
Additionally, experience in the ion trapping
community with micro-fabricated traps has shown that insulating structures
with feature sizes on the scale of 10 $\mu$m can satisfactorily
withstand $V_{max}=1000$ V, for example, Ref.~\cite{sterling2014fabrication}.
We thus speculate that kapton-based PCB carpets probably cannot safely
hold voltages much above 500 V RF at ${\cal O}(50\,\mu$m$)$ pitch
sizes, whereas nano-fabrication methods have been demonstrated which
allow up to 1000 V over 10 $\mu$m spacings, in principle. 

\medskip{}

\textbf{RF frequency $\Omega$:} In our calculations we have considered fixed
RF frequencies of 13.56 MHz due to easy availability of RF generators
at this frequency for industrial application in plasma production~\cite{regulations2012article}, though in principle, carpet operation
at any RF driving frequency is conceivable. In the heavily damped regime
all quantitative results concerning stability are nearly independent of RF frequency in
realistic ranges, so we do not scan over this parameter.  Since RF carpet power
consumption is minimized by choosing the lowest frequency possible, it is likely that such
a system would be operated at the lowest driving frequency where stable operation can be achieved.  
For the small pitches considered here this is around 1~MHz, below which the micro-radius 
will become comparable to the equilibrium ion height, 
leading to trajectories which are no longer well represented by macro-motion within the pseudo-potential well.

\medskip{}

\textbf{Push field $E_{push}$:} The push field is what drives ions
toward the carpet and, counteracting the RF levitating force, creates
the trapping effect. Normal operation of a time projection chamber
using high pressure xenon gas would naturally create a perpendicular
 field into the cathode of order $E_{push}=300$ V cm$^{-1}$. This is too
large for stable RF carpet transport under reasonable operating conditions,
since typical push fields in RF carpet systems are often of the order
10--30 V cm$^{-1}$.  As discussed in Sec.~\ref{sec:GeneralConsideration}, this
mis-match invokes the need for a gating grid which
switches when ions arrive. Using the scheme described earlier, a push field of 
20~V cm$^{-1}$ can be applied to allow for efficient ion transport.

\medskip{}

\textbf{Temperature} $T$: All the NEXT detectors to date have operated
at room temperature, $T\sim293$ K. Reduced temperature operation
of future phases has been considered in order to reduce the dark noise
present in silicon photo-multipliers to levels that may allow them
to be used for energy measurement in addition to tracking,
and to relieve mechanical constraints on the system by offering higher
densities at lower pressures. Reducing the operating temperature of
the detector will cause ions in the pseudo-potential to thermalize
at a lower temperature, narrowing their distribution around the stable
trajectory and leading to enhanced stability. This is expected to
offer a modest enhancement to RF carpet efficiency, since the mean
cluster sizes and ion mobilities are also temperature dependent. Xenon
freezes at around 219K at 10 bar, and so certainly the system cannot be operated
below this temperature. In what follows we will evaluate transport
distances at both room temperature and at $T=220$ K to illustrate
the (relatively modest) effects of temperature changes.

\begin{figure}
\begin{centering}
\includegraphics[width=0.99\columnwidth]{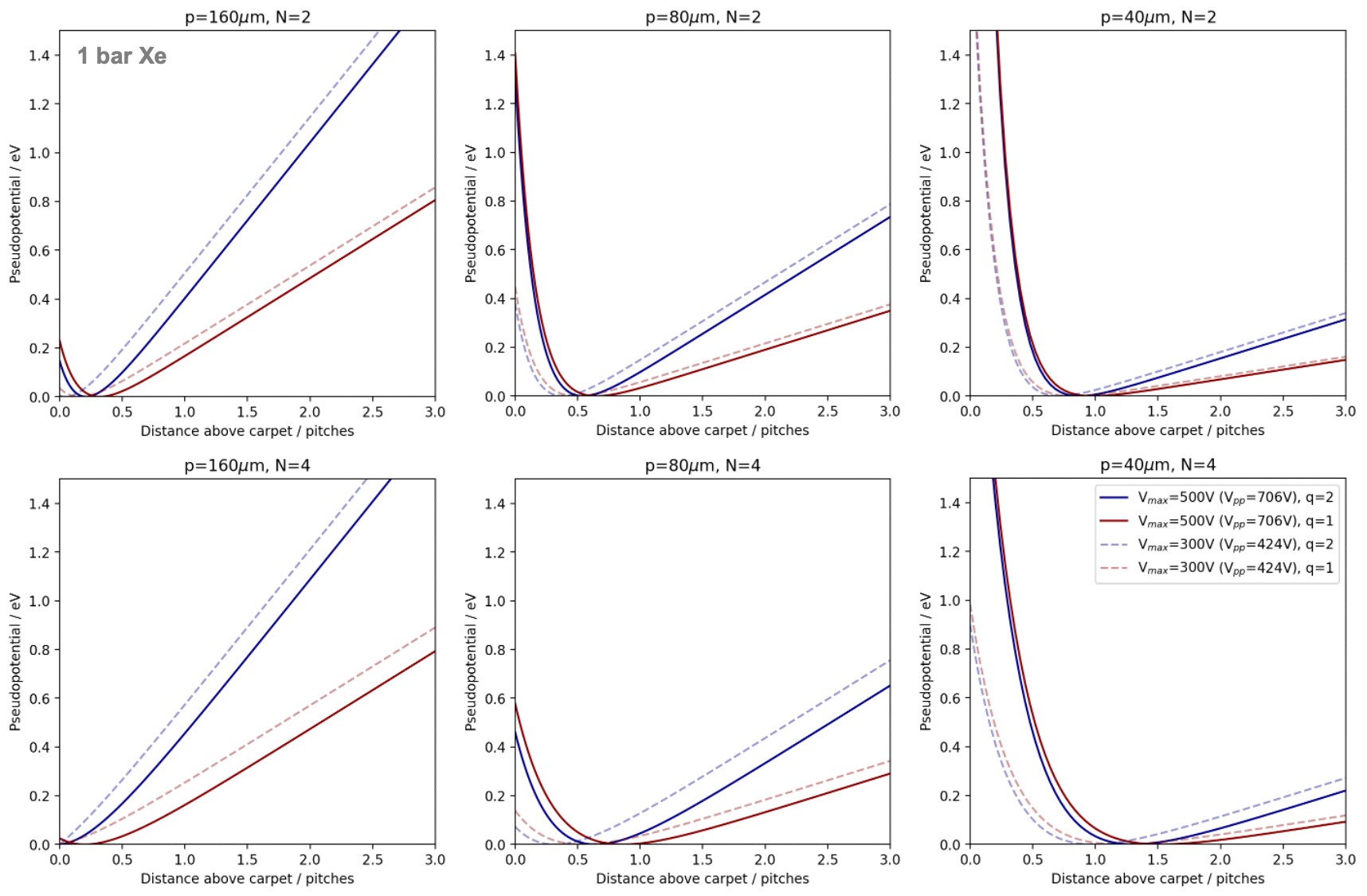}
\par\end{centering}
\caption{Pseudo-potential wells calculated for one bar xenon gas for singly
and doubly charged barium ions at room temperature. Stable trapping
is clearly possible for Ba$^{2+}$ in all but the least favorable
scenarios. \label{fig:OneBarPlots}}
\end{figure}

\begin{figure}
\begin{centering}
\includegraphics[width=0.5\columnwidth]{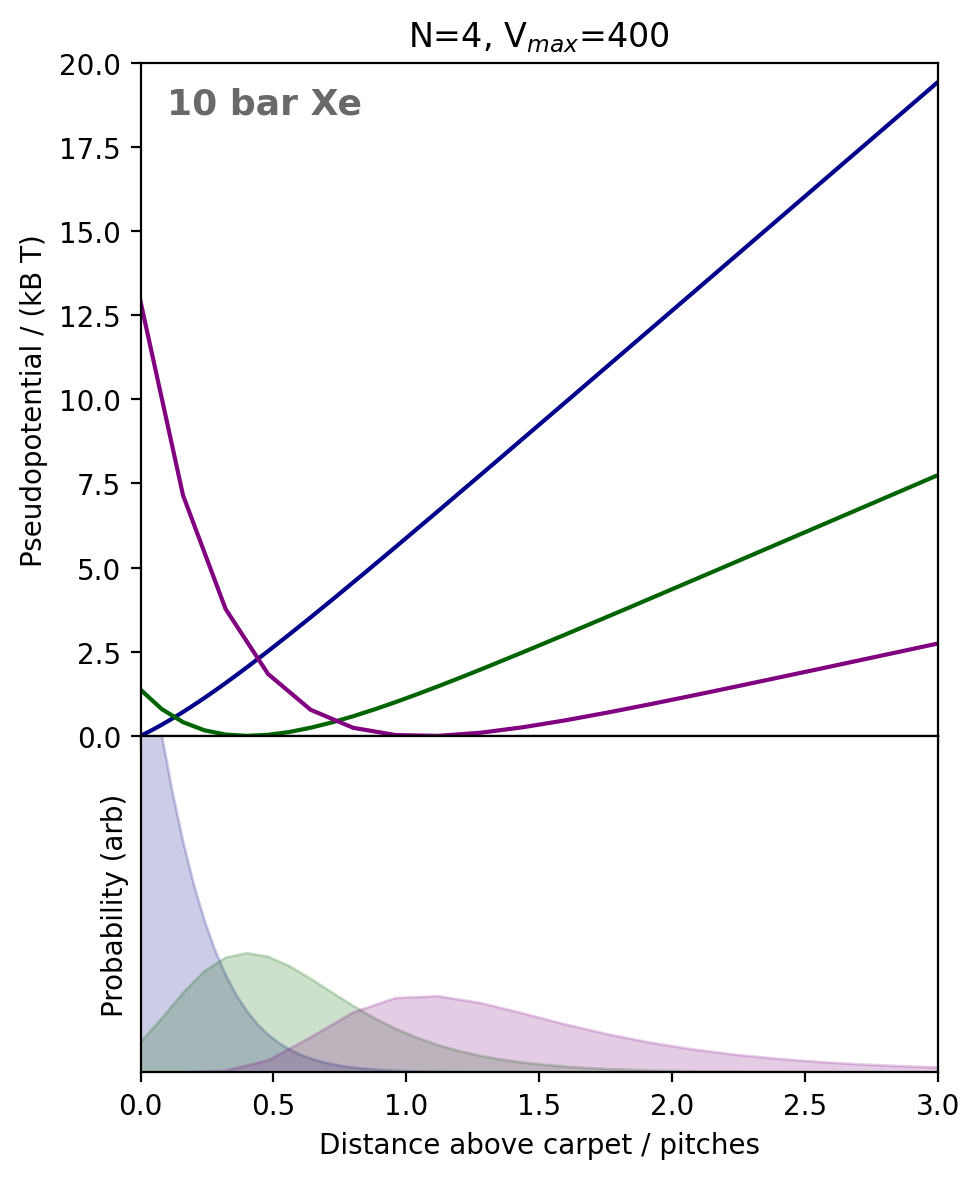}\includegraphics[width=0.5\columnwidth]{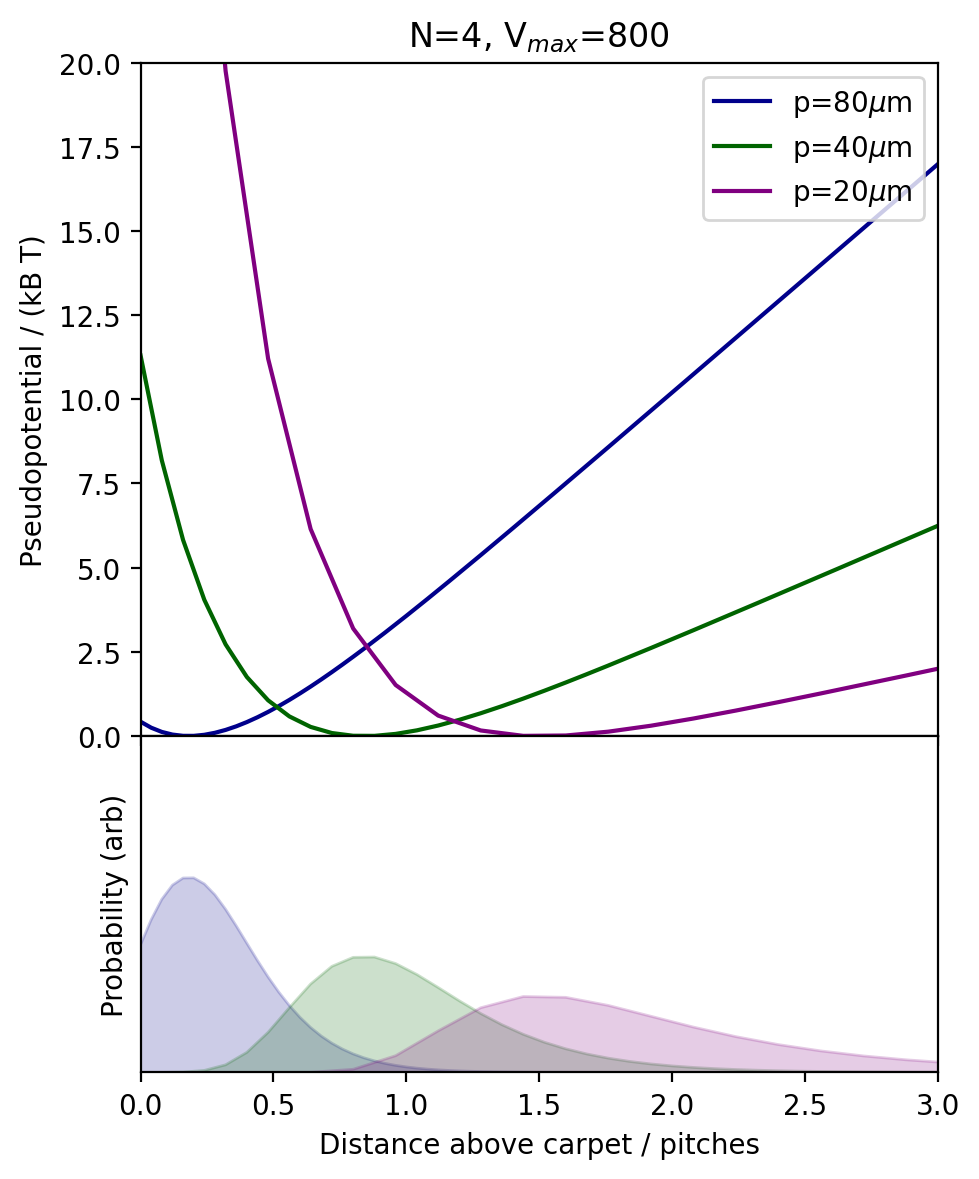}
\par\end{centering}

\caption{Examples of potential curves and ion height thermal probability distributions
in 10 bar xenon gas for Ba$^{2+}$  within its xenon cluster with N=4 RF carpets. The three
colored curves show three different pitches achievable with nano-fabrication
methods; the two figures show two different RF driving voltages.
In the upper panel we show the pseudo-potential in units of $k_B\,T$ and
in the lower, the equilibrium ion height distribution. \label{fig:Examples-of-potential10}}
\end{figure}

\section{Operating conditions for RF carpets in high pressure xenon gas\label{sec:Operating-conditions-for}}

We now apply the methods of the previous sections to the question
of ion transport efficiency in xenon gas at atmospheric pressure and
above. We begin by considering operation of RF carpets in xenon at
1 bar, which appears likely to be achievable with existing technical solutions
using kapton-based PCBs. Fig.~\ref{fig:OneBarPlots} shows the calculated
pseudo-potential for various pitches and RF voltages given a push
field of $E_{push}$=20 V cm$^{-1}$ and RF frequency of 13.56 MHz in one
bar of xenon gas. Potentials for both singly and doubly charged barium
ions are shown, and all calculations include the effects of clustering
on both mobility and molecular ion mass. At 80 $\mu$m pitch and
1 bar pressure, stable transport requires a voltage near to but not in excess of the edge of
viability for kapton substrates.  As expected,
reducing the carpet pitch leads to dramatic improvement in the depth
and vertical position of the pseudo-potential well. 

As the pressure is raised, stable transport becomes increasingly challenging.
Fig.~\ref{fig:Examples-of-potential10} shows pseudo-potential curves
as well as calculated thermal probability distributions at room temperature
for several RF carpet pitches and RF voltages at 10~bar. Consideration of the probability distributions 
shown in the lower panel of this figure illustrates
that stable transport cannot be expected at 10 bar for carpet pitches
above a few tens of microns, or electrode-to-electrode voltages below 700--800V. 
 At operating points with lower voltages or larger pitches,
the ion height probability distributions clearly have large tails penetrating the 
carpet surface, implying rapid ion losses.

The figure of merit that is the ultimate determinant of viability
for RF carpets for ion transport in xenon gas detectors is the mean
transverse ion survival distance. We define as our requirement here
as at least 15 cm predicted transport distance, though it is clear from
the curves of, for example, Fig.~\ref{fig:Comparison-of-simulated}
that once the threshold for stable transport is crossed, mean travel
distances climb extremely rapidly with voltage. The difference
in operating voltage required for a mean distance of 15 cm
or 50 cm is thus only at the few percent level. To establish the possible parameter space
of RF carpet operation we first fix the operating pitch, pressure,
push field, frequency, ion charge, mobility and temperature. We then calculate
the curve of mean survival distance vs. RF voltage using the methods
of Sec.~\ref{sec:Stochastic-effects-from}. Finally we interpolate
this curve to evaluate the voltage threshold where 15 cm mean transport
distance is achieved. This will be considered as the threshold voltage
for efficient transport at this operating parameter point.

\begin{figure}
\begin{centering}
\includegraphics[width=0.9\columnwidth]{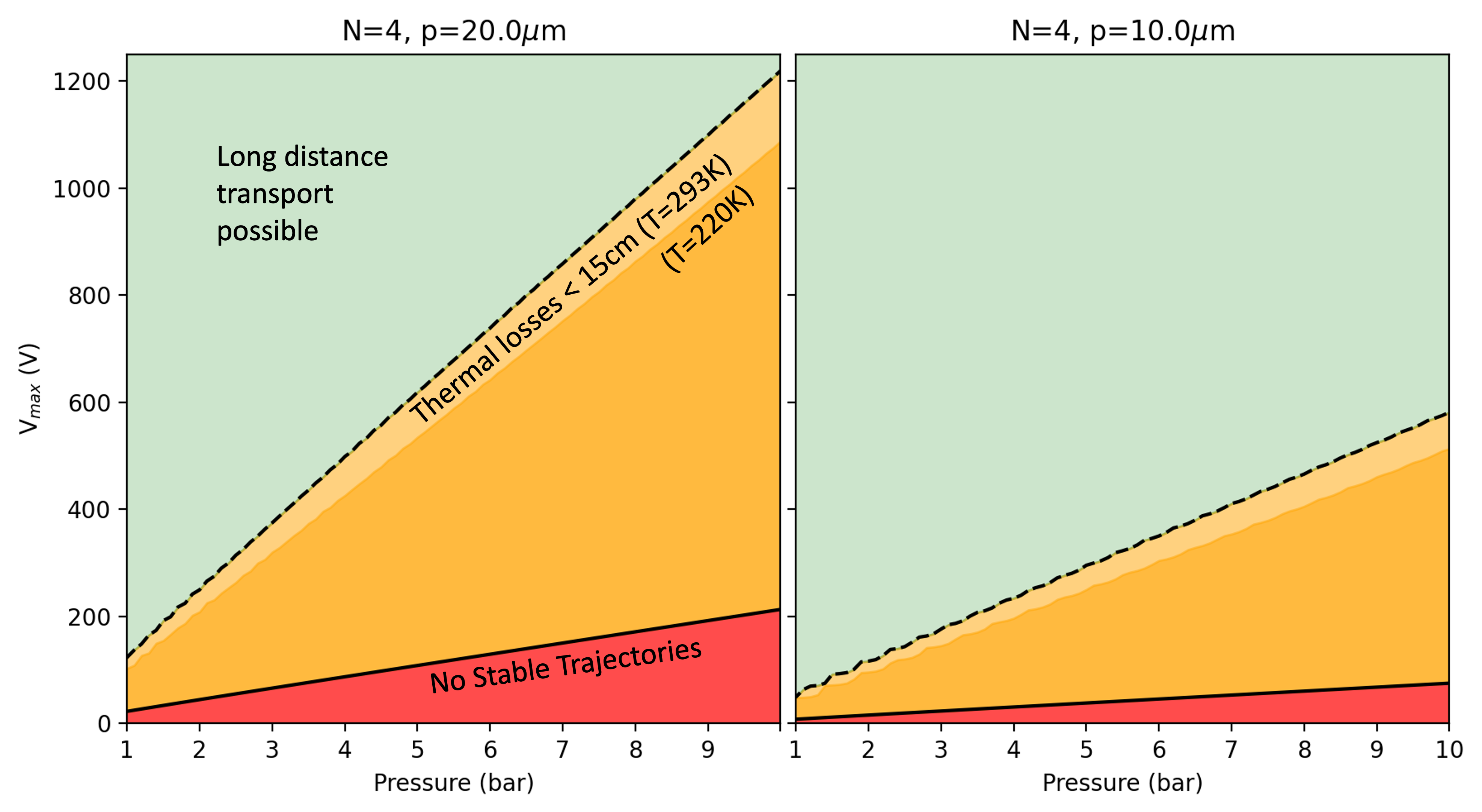}
\par\end{centering}
\caption{RF carpet stability regimes with fixed pitch and variable pressure.
In the red region, the minimum of the psuedo-potential is not above the carpet surface. In the yellow regions, ions travel less than 30~cm on average at the two tested temperatures. The green region covers the parameter values where ions travel more than 30~cm on average at 293~K.  Due to the rapid threshold-like form of the transport efficiency, the regions for 50~cm average travel are only different by a few percent in RF voltage. \label{fig:RF-carpet-stability}}
\end{figure}

\begin{figure}
\begin{centering}
\includegraphics[width=0.9\columnwidth]{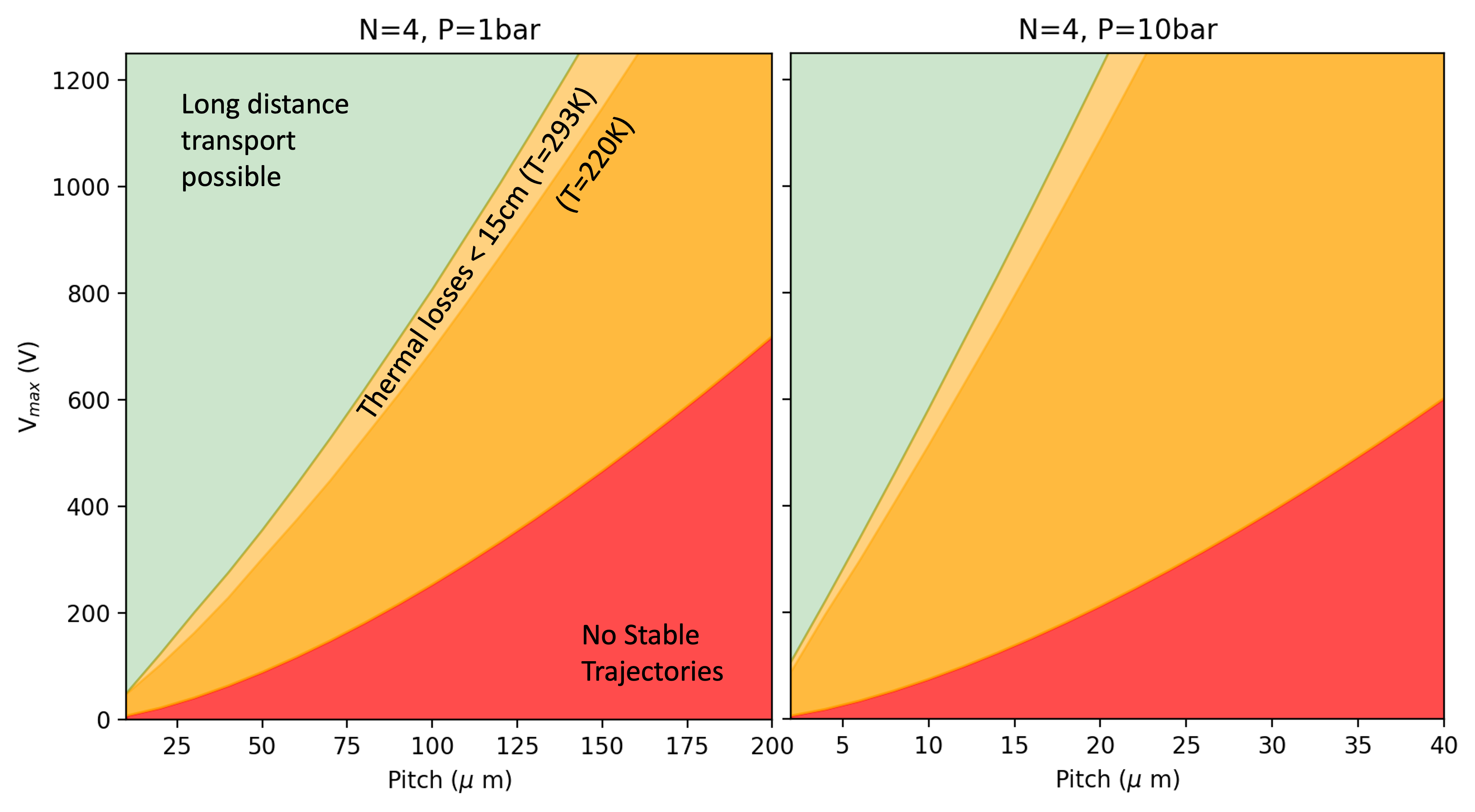}
\par\end{centering}
\caption{RF carpet stability regimes with fixed pressure and variable pitch.
See text for more details. In the red region, the minimum of the pseudo-potential is not above the carpet surface. In the yellow regions, ions travel less than 15~cm on average at the two tested temperatures. The green region covers the parameter values where ions travel more than 15~cm on average at 293~K\label{fig:RF-carpet-stability-1}.   Due to the rapid threshold-like form of the transport efficiency, the regions for 50~cm average travel are only different by a few percent in RF voltage.}
\end{figure}

Results of scans in pressure at fixed pitch are shown in Fig.~\ref{fig:RF-carpet-stability},
and complementary scans in pitch at fixed pressure are reported in
Fig.~\ref{fig:RF-carpet-stability-1}. These figures are the main
quantitative results of this work, encoding the operating requirements
for RF carpets for high pressure xenon gas. Also shown on the figure
is the region where even absent thermal fluctuations, no stable ion
trajectories exist. The large disparity between this region and the
stable long-distance transport regime highlights how vital the incorporation
of finite temperature / stochastic effects is to predicting the stability
of RF carpet operation, in contrast to much less strongly damped vacuum-based
trap systems.

\section{Discussion and Conclusions \label{sec:Conclusions}}

In this paper we have explored the theoretical operating
conditions for $N-$phased radio-frequency carpets in high pressure gases.
 The micro and macro-motion of ions on
phased RF arrays has been derived via generalization of the Dehmelt potential formalism,
with analytic expressions obtained for all important transport parameters.  In high pressure environments,
the analytic solution for the macro-motion is modified by stochastic fluctuations due to buffer gas interaction.
Diffusion of the particle trajectory from the minimum of the pseudo-potential into the carpet surface due to
Brownian motion is the dominant mechanism of ion loss in high buffer gas densities, and thus a proper understanding
of this deviation is required for understanding transport efficiency of RF carpets in dense gases.  Detailed
collision-by-collision simulations in SIMION were used to validate a new model for RF carpet performance
based on thermodynamics and kinetic theory, which are used to accurately describe
 both the equilibrium ion height distribution and the mean ion loss rate, respectively.  The validity of this model
 alleviates the need for computationally expensive simulations, which become untenable for high operating pressures and 
  long transport distances. 

Using this newly validated method for analytic modeling of RF carpet transport, the operating
conditions required for transport of barium ions in xenon gas at pressures of 1--10 bar have been evaluated. 
If RF carpets can be operated with high transport efficiency in this 
 regime it would enable ion collection from large volumes of xenon gas, which may prove to
 be a key ingredient in addressing the presently unsolved problem of ion concentration to sensors for barium tagging.

Our calculations show that stable ion motion over conventionally available RF carpets based on
single layer PCBs on kapton appears viable up to 1 atmosphere
of pressure.   At this operating pressure, commercially available PCB manufacturing technologies~\cite{Averatek} can 
be used to produce RF carpet structures with pitches of order 50~$\mu$m on 30 cm diameter scales which would
likely be able to support the required maximum driving voltage of $V_{max}\sim$350 V.  Construction of devices to operate at 10~bar of pressure, on the other hand, appears far more challenging.  A structure with at most 15 $\mu$m pitch supporting at least $V_{max}$=800 V would likely be required.  Such parameters
have been demonstrably achieved in past devices for quantum information
science applications using nano-fabrication at single-chip scales~\cite{sterling2014fabrication,brown2021materials,mehta2014ion}.  To our knowledge, however, no device with such a fine feature size has been produced on the tens-of-centimeter scale required for this application.  Since 300~mm is a standard wafer size in CMOS processing~\cite{CMOS300}, with 450~mm under development~\cite{CMOS300}, there is reason to be hopeful that such a device may be realizable with existing manufacturing techniques, though its demonstration will require considerable R\&D.  RF carpet devices operable in high pressures that may be realized in this way would likely find other important applications beyond double beta decay searches.

Experimental work on RF carpets for neutrinoless double beta decay detectors is currently underway within the NEXT collaboration. A near-term goal is to fabricate a 15 $\mu$m pitch etched metal-on-sapphire RF carpet at 5~cm scale and demonstrate its performance for ion
transport in xenon gas.  Success in this endeavor may illuminate a path toward RF carpets operable in new pressure regimes,  potentially providing a new method of ion concentration for barium daughter ion tagging following neutrinoless double beta decay.

\section*{Acknowledgements:}

We thank Ben Smithers and Jackie Baeza Rubio for careful proof-reading, and Yuan Mei for
thoughtful suggestions which were incorporated into the draft.  The University of Texas at Arlington NEXT group is supported by the Department of Energy under Early Career Award number DE-SC0019054 (BJPJ), by Department of Energy Award DE-SC0019223 (DRN), the National Science Foundation under award number NSF CHE 2004111 (FWF), and the Robert A Welch Foundation, Y-2031-20200401 (FWF).  The NEXT Collaboration acknowledges support from the following agencies and institutions: the European Research Council (ERC) under the Advanced Grant 339787-NEXT; the European Union's Framework Programme for Research and Innovation Horizon 2020 (2014--2020) under the Grant Agreements No.\ 674896, 690575 and 740055; the Ministerio de Econom\'ia y Competitividad and the Ministerio de Ciencia, Innovaci\'on y Universidades of Spain under grants FIS2014-53371-C04, RTI2018-095979, the Severo Ochoa Program grants SEV-2014-0398 and CEX2018-000867-S, and the Mar\'ia de Maeztu Program MDM-2016-0692; from Fundacion Bancaria la Caixa (ID 100010434), grant code LCF/BQ/PI19/11690012; the Generalitat Valenciana of Spain under grants PROMETEO/2016/120 and SEJI/2017/011; the Portuguese FCT under project PTDC/FIS-NUC/2525/2014 and under projects UID/FIS/04559/2020 to fund the activities of LIBPhys-UC; the Pazy Foundation (Israel) under grants 877040 and 877041; the US Department of Energy under contracts number DE-AC02-06CH11357 (Argonne National Laboratory), DE-AC02-07CH11359 (Fermi National Accelerator Laboratory), DE-FG02-13ER42020 (Texas A\&M). DGD acknowledges support from the Ram\'on y Cajal program (Spain) under contract number RYC-2015-18820. JM-A acknowledges support from Fundaci\'on Bancaria la Caixa (ID 100010434), grant code LCF/BQ/PI19/11690012, and from the Plan GenT program of the Generalitat Valenciana, grant code CIDEGENT/2019/049. Finally, we are grateful to the Laboratorio Subterr\'aneo de Canfranc for hosting and supporting the NEXT experiment.

\section*{Appendix: The Dehmelt potential\label{sec:Appendix:-The-Dehmelt}}
The levitating force in RF carpets operated with a purely periodic
field strength can be understood in terms of a Dehmelt potential,
the same basic principle operational in Paul traps. The Dehmelt potential
is an interesting and not totally intuitive effect, where an RF field
which averages to zero over a cycle everywhere at each point in space
can still exert an overall force on a particle when averaged over
time, if it is spatially inhomogeneous. Consider an electric field
with position $\vec{r}$ dependence of the form:

\begin{equation}
\vec{E}\left[\vec{r},t\right]=\vec{E}_{0}\left[\vec{r}\right]\cos\Omega t
\end{equation}

This field has a position dependence and fast time dependence at the
RF frequency $\Omega$. We consider three cases of particle motion.
First, a ballistic (vacuum-like) motion, and then a fully viscous
(high pressure gas-like) motion, and finally, something in between
(which it turns out is the most relevant case for operation RF carpets).

\medskip{}

\textbf{Fully Ballistic Case}: A particle moving in this field in
a ballistic, non-viscous setting obeys equation of motion:

\begin{equation}
m\ddot{r}=q\vec{E}\left[\vec{r}(t),t\right].\vec{r}=q\vec{E}_{0}\left[\vec{r}(t)\right]\cos\left(\Omega t\right).\label{eq:EOfM}
\end{equation}

We seek solutions defined by having a periodic ``micro-motion''
$\vec{\xi}$ on the RF timescale, superposed on a ``macro-motion''
$\vec{z}$ on slower timescales:

\begin{equation}
\vec{r}(t)=\vec{z}(t)+\vec{\xi}(t).
\end{equation}

Substituting this expression into Eq. \ref{eq:EOfM} we find:

\begin{equation}
m\ddot{\vec{z}}+m\ddot{\vec{\xi}}=q\vec{E}_{0}\left[\vec{z}+\vec{\xi}\right]\cos\left(\Omega t\right).\label{eq:EOfM-1}
\end{equation}

Two approximations allow us to proceed to solve for the micro-motion.
First, we approximate that the cycles defined by the micro-motion
take place over regions where the electric field does not vary sustantially,
so that $\vec{E}_{0}\left[\vec{z}+\vec{\xi}\right]\sim\vec{E}[\vec{z}]$.
Second, we approximate that the timescales associated with the micro-motion,
$\tau\sim1/\Omega$ are much faster than the timescales associated
with the macro-motion so $\ddot{z}_{i}\ll\ddot{\xi}_{i}$ for all
$i$. Then we obtain an equation for the micro-motion alone:

\begin{equation}
m\ddot{\vec{\xi}}=q\vec{E}_{0}\left[\vec{z}\right]\cos\left(\Omega t\right)\label{eq:EOfM-1-1}
\end{equation}

This has the periodic solution:

\begin{equation}
\vec{\xi}=\frac{q}{m\Omega^{2}}\vec{E}_{0}\left[\vec{z}\right]\cos\left(\Omega t\right)\label{eq:EOfM-1-1-1}
\end{equation}

To solve for the macro-motion, we return to the equation of motion
and expand $\vec{E}[\vec{r}]$ to first order in $\vec{\xi}$:

\begin{equation}
m\ddot{\vec{z}}+q\vec{E}_{0}\left[\vec{z}\right]\cos\left(\Omega t\right)=q\left(1+\vec{\xi}(t)\cdot\nabla\right)\vec{E}_{0}\left[\vec{z}\right]\cos\left(\Omega t\right)\label{eq:EOfM-1-2}
\end{equation}

Substituting the solution for the micro-motion:

\begin{equation}
m\ddot{\vec{z}}=\frac{q^{2}}{m\Omega^{2}}\left(\vec{E}_{0}\cdot\nabla\right)\vec{E}_{0}\cos^{2}\left(\Omega t\right)\label{eq:EOfM-1-2-2}
\end{equation}

Averaging over a period in $\Omega t$ to find the slow-timescale
behavior:

\begin{equation}
\langle\ddot{\vec{z}}\rangle=\frac{q^{2}}{m\Omega^{2}}\left\langle \left(\vec{E}_{0}\cdot\nabla\right)\vec{E}_{0}\cos^{2}\left(\Omega t\right)\right\rangle =\frac{1}{2}\frac{q^{2}}{m^{2}\Omega^{2}}\left(\vec{E}_{0}\cdot\nabla\right)\vec{E}_{0}\label{eq:EOfM-1-2-2-1}
\end{equation}

If the E field is homogenous then $\nabla\vec{E}=0$ and there is
no macro-motion. However, an interesting situation is apparent when
$\nabla\vec{E}\neq0$. Now there is a non-zero force, after time averaging,
even though the RF field at each point in space averages to zero. 

\begin{equation}
\langle\ddot{\vec{z}}\rangle=\frac{q^{2}}{4m^{2}\Omega^{2}}\nabla\left(E_{0}^{2}\left[\vec{z}\right]\right)\label{eq:EOfM-1-2-2-1-1}
\end{equation}

This can be compactly written:

\begin{equation}
\langle\ddot{\vec{z}}\rangle=q\nabla V
\end{equation}

With V the effective potential:

\begin{equation}
V=\frac{qE_{0}^{2}[\vec{z}]}{4m^{2}\Omega^{2}}
\end{equation}

It is this repulsive potential that acts to levitate ions over any
RF structures that have non-uniform and time varying electric fields
near their surfaces. 

\medskip{}

\textbf{Fully Viscous Case}: We can also consider a scenario with
a viscous motion, as we would expect in very high pressure environments.
We now take as starting point the viscous equation of motion:

\begin{equation}
\dot{\vec{r}}=q\mu\vec{E}_{0}\left[\vec{r}(t)\right]\cos\left(\Omega t\right)\label{eq:EOfM-2}
\end{equation}

Here $\mu$ is the mobility and the dynamical term on the left involves
only one time derivative. Again we seek solutions defined by a micro-motion
$\vec{\xi}$ on the RF timescale superposed on a macro-motion $\vec{z}$
on slower timescales. Solving for the micro-motion using the same
approximations as before yields:

\begin{equation}
\vec{\xi}(t)=\frac{\mu q}{\Omega}\vec{E}_{0}\left[\vec{z}\right]\sin\left(\Omega t\right)\label{eq:EOfM-1-1-1-1}
\end{equation}

And so:

\begin{equation}
\dot{\vec{z}}=\frac{\mu^{2}q^{2}}{\Omega}\left(\vec{E}_{0}\cdot\nabla\right)\vec{E}_{0}\frac{1}{2}\sin\left(2\Omega t\right)
\end{equation}

In this case, however, when time averaging over an oscillation we
find there is no longer a levitating force:

\begin{equation}
\langle\dot{\vec{z}}\rangle=0
\end{equation}

Thus is a scenario where the ion motion is trully viscous over the
RF timescale, ion levitation using RF fields is not possible.

\medskip{}

\textbf{Intermediate case: }It would be an over-simplification to
state that ions exhibit viscous motion in high pressure gases but
ballistic motion in low pressure ones. When we consider ion swarms
propelled by DC drift fields, it is understood that the particles
are subject to viscous motion on the long timescales required to cross
a particle detector. However, they still accelerate ballistically
between collisions with gas atoms. The ion motion on a microscopic
scale thus exhibits both ballistic and viscous elements. 

The full equation of motion in the presence of both ballistic and
viscous terms can be written as:

\begin{equation}
m\ddot{\vec{r}}+\frac{q}{\mu}\dot{\vec{r}}=q\vec{E}_{0}\left[\vec{r}(t)\right]\cos\left(\Omega t\right)\label{eq:EOfM-3}
\end{equation}

Following the same approach as before we can solve for the fast micro-motion:

\begin{equation}
m\Omega^{2}\ddot{\vec{\xi}}+\frac{q}{\mu}\Omega\dot{\vec{\xi}}=q\vec{E}_{0}\left[\vec{r}(t)\right]\cos\left(\Omega t\right)\label{eq:EOfM-3-1}
\end{equation}

The solution to this second order differential equation is a damped,
periodic motion with a phase shift:

\begin{equation}
\vec{\xi}=\frac{q}{m\Omega}\frac{\vec{E}_{0}}{\sqrt{\Omega^{2}+D^{2}}}\cos(\Omega t+\beta)\quad\quad\tan\beta=\frac{D^{2}}{\Omega^{2}}
\end{equation}

Above we have introduced the ``Damping factor'':

\begin{equation}
D=\frac{e}{m\mu}
\end{equation}

\begin{figure}
\begin{centering}
\includegraphics[width=0.6\columnwidth]{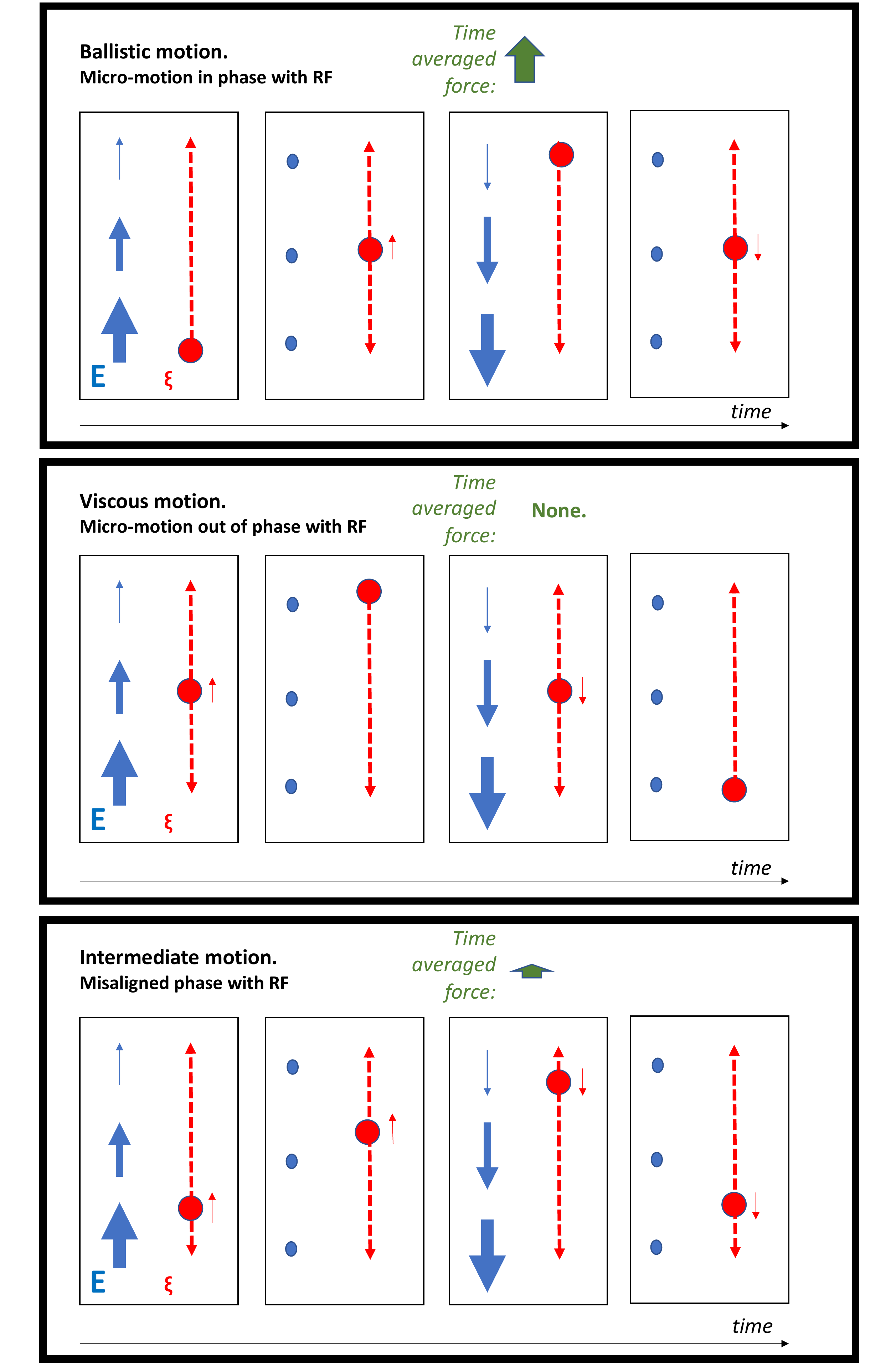}
\par\end{centering}
\caption{Illustration of the reinforcing effect of the non-uniform, time-dependent
field that gives rise to the Dehmelt potential in ballistic and semi-ballistic
but not in viscous motion. \label{fig:Illustration-of-the}}
\end{figure}

The ballistic limit corresponds to $\tan\beta\rightarrow0$ so $\beta\rightarrow0$
and the micro-motion totally in phase with the RF; whereas the viscous
limit corresponds to $\tan\beta\rightarrow\infty$ so $\beta\rightarrow\pi$
and the micro-motion totally out of phase with the RF. This principle
can be used to understand where the effective force from the Dehmelt
potential comes from, by considering the reinforcing action (or lack
of) of the non-uniform electric field at different points in the micro-cycle.
Fig. \ref{fig:Illustration-of-the} illustrates this principle. 

Proceeding with a derivation of the macro-motion:

\begin{equation}
m\langle\ddot{\vec{z}}\rangle+\frac{q}{\mu}\langle\dot{\vec{z}}\rangle=\frac{q}{4m\Omega}\frac{1}{\sqrt{\Omega^{2}+D^{2}}}\left(\vec{E}_{0}\cdot\nabla\right)\vec{E}_{0}\left\langle \cos(\Omega t+\beta)\cos\left(\Omega t\right)\right\rangle \label{eq:EOfM-1-2-1-1}
\end{equation}

A trigonometric identity helps us with the time average on the right,
and only the first term survives being time-averaged:

\begin{equation}
\left\langle \cos(\Omega t+\beta)\cos\left(\Omega t\right)\right\rangle =\left\langle \left(\cos(\Omega t)\cos(\beta)-\sin(\Omega t)\sin(\beta)\right)\cos\left(\Omega t\right)\right\rangle \label{eq:EOfM-1-2-1-1-1}
\end{equation}

\begin{equation}
=\frac{1}{2}\cos\beta.
\end{equation}

We can also write this as:

\begin{equation}
\cos\beta=\frac{1}{\sqrt{1+\tan^{2}\theta}}=\frac{\Omega}{\sqrt{\Omega^{2}+D^{2}}}
\end{equation}

Thus our equation of motion has become:

\begin{equation}
m\langle\ddot{\vec{z}}\rangle+\frac{q}{\mu}\langle\dot{\vec{z}}\rangle=\frac{q}{4m}\frac{E_{0}}{\Omega^{2}+D^{2}}\nabla\left(E_{0}^{2}\left[\vec{z}\right]\right)
\end{equation}

The statement that on timescales associated with macro-motion the
dynamics are viscous corresponds to the condition $\frac{q}{\mu}\left\langle \dot{z}\right\rangle \gg m\left\langle \ddot{z}\right\rangle $.
In this scenario we find:

\begin{equation}
\frac{q}{\mu}\langle\dot{\vec{z}}\rangle=\nabla V
\end{equation}

\begin{equation}
V=\frac{q}{4m}\frac{1}{\Omega^{2}+D^{2}}E_{0}^{2}\left[\vec{z}\right]
\end{equation}

As required. We see then that interestingly, even in the extremely
viscous limit where $D\gg\Omega$, a Dehmelt potential remains active,
though suppressed by the damping factor. 

 \bibliographystyle{elsarticle-num} 
 \bibliography{rfbib}
\end{document}